\documentclass[journal]{IEEEtran}

\newcommand{\T}{{\top}}

\usepackage{amssymb}
\usepackage[english]{babel}
\usepackage[utf8x]{inputenc}
\usepackage{amsmath}
\usepackage[overload]{empheq}
\usepackage{framed}
\usepackage{marginnote}

\usepackage{caption}

\usepackage{lipsum}

\usepackage{mathtools}
\usepackage{cuted}

\usepackage{stfloats}

\usepackage{todonotes}
\usepackage{multirow,booktabs}
\usepackage{graphicx}
\usepackage{amsfonts}
\graphicspath{{images/}}
\usepackage{subfigure}
\usepackage[justification=centering]{caption}
\usepackage{epstopdf}
\usepackage{gensymb}
\usepackage[flushleft]{threeparttable}
\usepackage{balance} 
\usepackage{bm}
\usepackage{cite}

\usepackage{float}

\usepackage{algorithmic}
\usepackage{algorithm}

\usepackage{enumerate}

\allowdisplaybreaks

\newtheorem{remark}{Remark}

\begin{document}


\title{Real-Time Optimal Lithium-Ion Battery Charging Based on Explicit Model Predictive Control}

\author{
	\vskip 1em
	Ning Tian, \emph{Student Member}, \emph{IEEE},
	Huazhen Fang, \emph{Member}, \emph{IEEE}, and Yebin Wang, \emph{Senior Member}, \emph{IEEE}

	\thanks{
This work was supported in part by the National Science Foundation under Awards CMMI-1763093 and CMMI-1847651. Paper no. TII-20-0598. (\emph{Corresponding author: Huazhen Fang.})

	Ning Tian and Huazhen Fang are with the Department of Mechanical Engineering, University of Kansas, Lawrence, KS 66045, USA (e-mail: ning.tian@ku.edu, fang@ku.edu). 

Yebin Wang is with the Mitsubishi Electric Research Laboratories, Cambridge,
MA 02139, USA (e-mail: yebinwang@ieee.org).
		
	}
}

\maketitle

\begin{abstract}
The rapidly growing use of lithium-ion batteries across various industries highlights the pressing issue of optimal charging control, as charging plays a crucial role in the health, safety and life of batteries. The literature increasingly adopts model predictive control (MPC) to address this issue, taking advantage of its capability of performing optimization under constraints. However, the computationally complex online constrained optimization intrinsic to MPC often hinders real-time implementation. This paper is thus proposed to develop a framework for real-time charging control based on explicit MPC (eMPC), exploiting its advantage in characterizing an explicit solution to an MPC problem, to enable real-time charging control. The study begins with the formulation of MPC charging based on a nonlinear equivalent circuit model. Then, multi-segment linearization is conducted to the original model, and applying the eMPC design to the obtained linear models leads to a charging control algorithm. The proposed algorithm shifts the constrained optimization to offline by precomputing explicit solutions to the charging problem and expressing the charging law as piecewise affine functions. This drastically reduces not only the online computational costs in the control run but also the difficulty of coding. Extensive numerical simulation and experimental results verify the effectiveness of the proposed eMPC charging control framework and algorithm. The research results can potentially meet the needs for real-time battery management running on embedded hardware. 

\end{abstract}




\section{Introduction}\label{Sec:Introduction}

Lithium-ion batteries (LiBs) have seen ever-increasing application across various sectors, including consumer electronics, electrified transportation and renewable energy, due to their appealing features like high voltage, high energy and power density, no memory effect, low self-discharge rates and long service life~\cite{Wang:CSM:2018}. This trend has been driving a surge of research on advanced battery management to ensure the performance, safety and longevity of LiBs. Among the problems of interest, a critical one is optimal charging design in pursuit of two main objectives: reducing side reactions and effects to prolong life, and increasing the charging speed to meet efficiency needs. This work proposes a novel optimal charging control approach based on the notion of model predictive control (MPC). While accommodating the above objectives, it is particularly designed {\em via} exploiting the recent advances of explicit MPC (eMPC) to attain real-time implementation. 
The proposed approach may find important prospective use in future real-time battery management systems.

\subsection{Literature Review}

Finding the best ways to charge LiBs has attracted sustained attention in the past two decades. Currently, the most popular industrial practice is the so-called constant-current/constant-voltage (CC/CV) charging~\cite{hussein2011review}. It applies a constant current to charge a LiB cell until it reaches a threshold voltage and then enforces a constant voltage to charge the cell at a gradually diminishing current. Another often endorsed practice is pulse charging that feeds energy into a battery using current pulses~\cite{cope1999art}. 
These methods, however, usually involve some heuristic determination of charging parameters, giving only empirical or conservative guarantee for charging safety and speed. This hence has motivated researchers to develop optimal charging protocols by combining physics-based LiB models and optimization to meet certain objectives concerning LiB health and/or charging time. A study is offered in~\cite{Suthar:PCCP:2014} to build current profiles that can maximize the charge stored in a given time while suppressing the internal stress buildup, using a single particle model (SPM) supplemented with an intercalation-induced stress generation model. To enhance the conventional pulse charging, the study in~\cite{Fang:JES:2018} optimizes the magnitudes and duty cycles of current pulses to reconcile health effects with rates of charging. 
The investigations in~\cite{Hu:JPS:2013,liu2016battery,Perez:TVT:2017,liu2018charging} lead to the design of health-aware, fast and thermal-safe charging protocols {\em via} multi-objective optimization based on coupled electro-thermal-aging models.

It is known that charging protocols are first generated offline and then run online, thus subjecting LiB charging to de facto open-loop control. Nonetheless, closed-loop control is arguably more capable of improving the charging performance, since it incorporates dynamically the feedback about a LiB's present state to regulate the charging process. The past years have witnessed a growing body of work on this subject. Linear quadratic control is leveraged in~\cite{Fang:TCST:2017} to enable health-aware LiB charging, with cost functions therein by design restricting the use of aggressive currents. Meanwhile, MPC, a constrained optimal control strategy, holds considerable promise here for two reasons. 1) It can handle hard state and input constraints. This gives a leverage to guarantee satisfaction of health- or safety-related constraints necessary for LiB operation. 2) It can optimize different kinds of objective functions to meet different charging needs or considerations. 
As another benefit, its formulation well admits nonlinear systems, thus bearing applicability to different types of nonlinear LiB models.

A lead is taken in~\cite{Klein:ACC:2011} with the development of minimum-time charging control by applying nonlinear MPC to a 1-D electrochemical model of LiBs. However, a barrier in the way of MPC-based charging is the high computational complexity that results from the numerical constrained optimization procedure at the core of an MPC algorithm. This can be more serious in the context of complex models, e.g., nonlinear electrochemical models involving various partial differential equations. Significant research hence has been devoted to computationally efficient MPC charging control design. The study in~\cite{Liu:DSMC:2016} considers nonlinear MPC for SPM and exploits the differential flatness of Fick's law of diffusion to reduce computational load. As another important way, model reduction is often used in the literature to simplify an electrochemical model and make it amenable for the design of efficient MPC. For example, the approach in~\cite{Zou:Mechatronics:2018} linearizes a nonlinear electrochemical model successively along a reference SoC trajectory. Other examples, e.g.,~\cite{Torchio:ACC:2015}, develop input-output approximations of a pseudo 2-D (P2D) model such as step response models and autoregressive exogenous models, so that application of MPC to them causes less computation. Particularly, the fast quadratic dynamic matrix control is used in~\cite{Torchio:ACC:2015} to further improve the computational efficiency.



Equivalent circuit models (ECMs) represent another appealing choice for MPC-based charging control due to their much less computation than electrochemical models. An early study is in~\cite{Yan:Energies:2011}, which yet adopts a genetic algorithm as the optimizer despite its costly computation. There is a consensus today that it is still a critical need to design fast MPC for ECMs for the sake of practical implementation. To this end, the literature derives either simpler models or computationally frugal control frameworks. The method in~\cite{Liu:JPS:2017} proposes to identify a time-series model recursively as an input-output approximation of the Thevenin model and takes advantage of its simplicity to achieve efficient generalized predictive control. Optimal charging based on the Thevenin model is formulated as a standard linear MPC problem in~\cite{Xavier:JPS:2015} that eases computation. Similarly, a linear-time-varying MPC method is proposed in~\cite{Zou:Energy:2017}. A hierarchical MPC design in~\cite{Ouyang:TII:2018} features the generation of reference current profiles at a slow time scale and the reference tracking at a faster time scale, which lowers the cost of computation.

\subsection{Research Overview and Contributions}

The above survey highlights the main advances in the development of computationally efficient MPC charging control. However, the existing methods still demand a relatively large amount of online computation, arising from the need to solve a constrained optimization problem at each sampling time. In addition, such online optimizers require strong computing capability, which is rarely available for the hardware on which a battery management system runs. 
A central challenge then lies in how to offload the primary part of the computational effort offline and run a lean controller online. The eMPC strategy, pioneered in~\cite{bemporad2002explicit,bemporad2015multiparametric}, is set to address this challenge. It pre-computes the control law offline by deriving explicit solutions to an MPC problem. The control law is composed of piecewise affine (PWA) functions of the system's current state, which can be run online through only straightforward arithmetic operations. The advantages of eMPC are remarkable. First, it can achieve MPC functionality with microsecond-millisecond online computational efficiency. Second, it is easy to code and executable on cheap embedded control hardware. Therefore, eMPC can hopefully provide a solution to bridge the gaps in computation and execution facing the current breed of MPC charging methods.

The contribution of this paper is threefold. 1) This work presents the first framework for eMPC-based optimal charging control. Distinguished from the literature, this framework exploits eMPC to considerably reduce the online computational time and complexity of coding, paving the way for real-time execution of charging control. 2) Based on the framework, an eMPC-based charging control algorithm is developed. The study considers the nonlinear double-capacitor (NDC) model, an ECM proposed in~\cite{Tian:IECON:2018}, and formulates a general nonlinear MPC charging problem. To deal with the nonlinearity inherent in the dynamics of batteries, it then simplifies the nonlinear MPC problem into a combination of approximate linear MPC problems. On such a basis, synthesis of eMPC-based control is performed to build an optimal charging control algorithm, which constructs on a set of PWA functions over a parameter space. 2) The proposed charging control framework and algorithm are evaluated through extensive simulations and experiments, with their performance well validated. 

\subsection{Organization}

The paper is organized as follows. Section~\ref{Sec:Modeling} introduces the NDC model along with the charging-related constraints. Section~\ref{Sec:Charging-Strategy} contains: 1) the statement of an MPC-based health-aware charging control problem, 2) the piecewise linearization of the model, and 3) the formulation of eMPC-based charging control. The proposed charging control law is evaluated by simulation in Section~\ref{Sec:Numerical-Simulation} and experiments in Section~\ref{Sec:Experimental-Validation}. Finally, Section~\ref{Sec:Conclusion} gathers concluding remarks and discussions of future research.

\section{The NDC Model for LiBs}\label{Sec:Modeling}

Laying the groundwork for the charging control design, this section describes the NDC model for LiBs and its governing dynamics. Further, the constraints to be enforced during charging are outlined.

\subsection{Model Description}

\begin{figure}[t]
\centering
\includegraphics[trim = {93mm 70mm 100mm 48mm}, clip, width=0.37\textwidth]{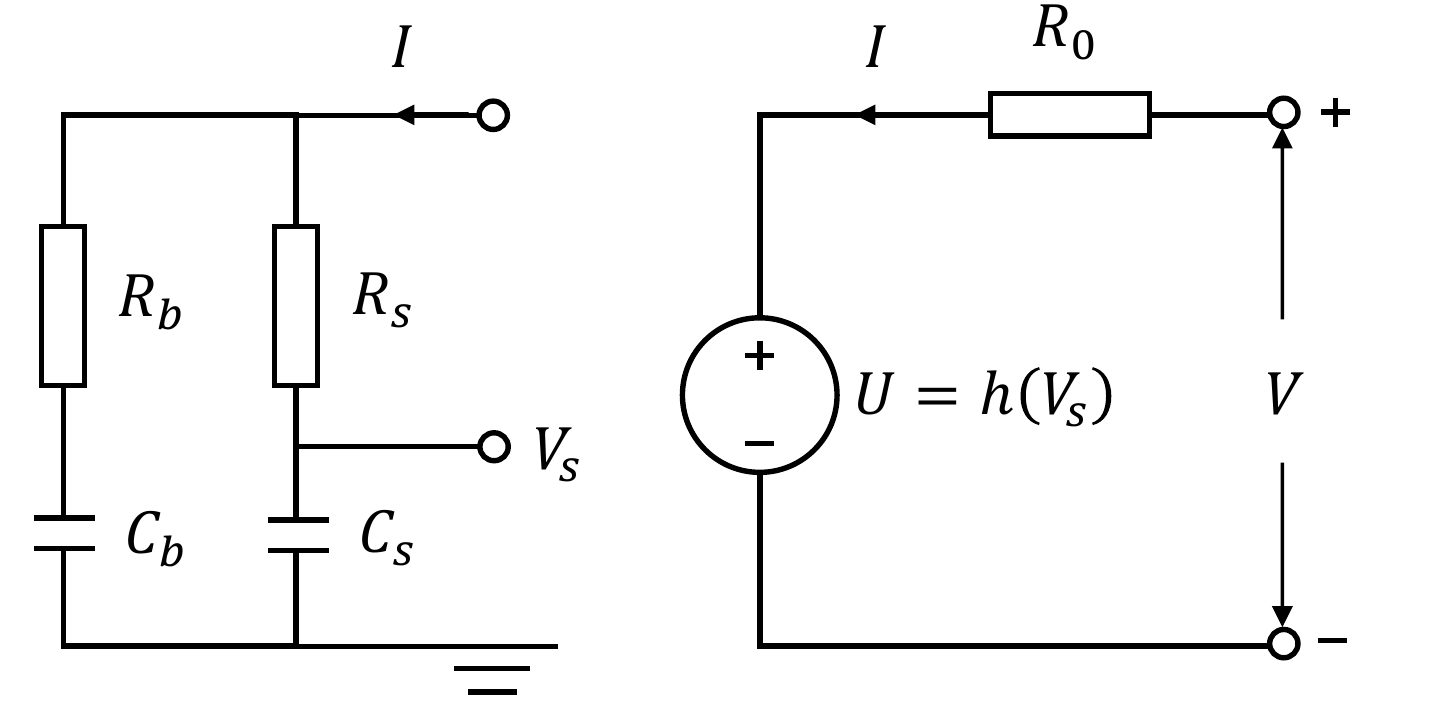}
\caption{The nonlinear double-capacitor model.} 
\label{Fig:NDC-model}
\vspace{-4mm}
\end{figure}

Developed in~\cite{Tian:IECON:2018}, the NDC model is shown in Figure~\ref{Fig:NDC-model}, which includes two main parts. The first part (left) includes two capacitors in parallel, $C_b$ and $C_s$, each serially connected with a resistor, $R_b$ and $R_s$, respectively. The two capacitors play the role of an electrode, providing storage for electric charge. The parallel connection between them allows the distribution and migration of charge within the electrode to be simulated. To be specific, the $R_s$-$C_s$ circuit conceptually corresponds to the electrode's surface region; 
the $R_b$-$C_b$ circuit represents the electrode's bulk inner part analogously. It generally holds that $C_b \gg C_s$ and that $R_b \gg R_s$. As is seen, $C_b$ is where the majority of the charge is stored, and $R_b$-$C_b$ underlies the low-frequency part of the charging/discharging response. By contrast, $C_s$ has a capacity much lower than $C_b$, and its voltage changes much faster during charging/discharging. This makes the $R_s$-$C_s$ circuit responsible for the high-frequency response. 
The second part (right) has a voltage source $U = h(V_s)$, where $V_s$ is the voltage across $C_s$. Hence, $U$ is similar to an open-circuit voltage (OCV) but depends on $V_s$ rather than SoC in a conventional sense. 
The last component is $R_0$, which is the internal resistance. From this overview, it is seen that the NDC model can simulate not only the diffusion of charge inside a LiB's electrode but also the nonlinear voltage behavior, making it distinguished from existing ECMs for LiBs.

The NDC model's dynamics is governed by the following state-space equations: 
\begin{subequations}\label{state-space-equation}
\begin{align}[left = \empheqlbrace\,] \label{state-equation}
\begin{bmatrix}
\dot{V}_{b}(t) \\ \dot{V}_s(t)
\end{bmatrix}
& =
A
\begin{bmatrix}
V_b(t) \\ V_s(t)
\end{bmatrix}
+
B
I(t), \\ \label{output-equation}
V(t) & = h(V_s(t))+R_0(V_s(t))I(t),
\end{align}
\end{subequations}
where $V_b$ is the voltage across $C_b$, $I$ the applied current with $I>0$ for charging and $I<0$ for discharging, and 
\begin{align*}
A = \begin{bmatrix}
-\frac{1}{C_b(R_b+R_s)} & \frac{1}{C_b(R_b+R_s)} \\
\frac{1}{C_s(R_b+R_s)} & -\frac{1}{C_s(R_b+R_s)}
\end{bmatrix}, \ 
B = \begin{bmatrix}
\frac{R_s}{C_b(R_b+R_s)}\\ \frac{R_b}{C_s(R_b+R_s)} 
\end{bmatrix}. 
\end{align*} 
Besides, $h(V_s)$ is parameterized as a polynomial. Experience suggests that a fifth-order polynomial can offer sufficient accuracy:
\begin{align*}
h(V_s) = \sum_{i=0}^5 \alpha_i V_s^i, 
\end{align*}
where $\alpha_i$ for $i=0,1,\ldots,5$ are coefficients. 

Here, $V_b$ and $V_s$ should be limited to a range. For both, the lower bound of the range is set to be $V_{s,\rm min}=0$ V, and the upper bound $V_{s,\rm max} = 1$ V for simplicity and without loss of generality. In other words, $V_b=V_s = 0$ V when the LiB is fully depleted ($\mathrm{ SoC} = 0\%$), and $V_b=V_s = 1$ V when it is fully charged ($\mathrm{ SoC} = 100\%$). Furthermore, the following shows the relation between $\mathrm{SoC}$ and $V_b$ and $V_s$:
\begin{align}\nonumber
{\rm SoC}&=\frac{C_b V_b+C_s V_s}
{(C_b +C_s)V_{s,\rm max} }\times 100\%,
\end{align}
where $(C_b +C_s)V_{s,\rm max} $ is a battery's total capacity, and $(C_b V_b+C_s V_s)$ the available capacity. Note that the function $h(\cdot)$ also characterizes the SoC-OCV relationship, because $V_b=V_s = {\rm SoC}$ when the battery is at equilibrium.

Finally, $R_0$ is also considered to monotonically increase with $V_s$. Such a dependence is described as 
\begin{align*}
R_0(V_s)=\beta_1+\beta_2 e^{-\beta_3(1-V_s)},
\end{align*}
where $\beta_i>0$ for $i=1, 2, 3$.

\subsection{Constraints}

To ensure health-conscious and safe charging, some constraints must be imposed during a charging process, a summary of which is as below. To begin with, the SoC must be constrained to avoid overcharging. That is, 
\begin{align}\label{inequality-SoC}
\rm{SoC_{min}} \leq \mathrm{SoC} \leq SoC_{max}.
\end{align}
The charging current and terminal voltage must also be subject to limitations to circumvent safety issues, implying
\begin{align}\label{inequality-I}
I_{\rm{min}} & \leq I \leq I_{\rm{max}}, \\ \label{inequality-V}
V_{\rm{min}} & \leq V \leq V_{\rm{max}}.
\end{align}

In addition, $V_b$ and $V_s$ should be kept within the pre-set range. Given the dynamics shown in~\eqref{state-equation}, $V_s\geq V_b$ always holds during charging if the LiB is at equilibrium initially. One hence only needs to limit $V_s$ by
\begin{align}\label{inequality-Vs}
V_{s,\mathrm{min}} \leq V_s \leq V_{s,\mathrm{max}}.
\end{align}
As explained in~\cite{Tian:IECON:2018}, $V_s$ strikes an analogy to the Li-ion concentration at the surface region of an electrode, which needs to be constrained as suggested in some studies about MPC charging based on electrochemical models, e.g.,~\cite{Zou:Mechatronics:2018}. This makes~\eqref{inequality-Vs} corresponding to those constraints though it is an ECM considered here.

The final constraint to add concerns $V_s-V_b$. It is seen that $V_s-V_b$ drives the migration of charge from $C_s$ to $C_b$ during charging. The study in~\cite{Tian:IECON:2018} points out that this variable is comparable to the Li-ion concentration gradients within an electrode when proving the approximate equivalence between the NDC model and the SPM. The Li-ion concentration gradients are a cause for internal stress buildup, and it can also lead to heating and formation of solid electrolyte interphase (SEI) film indirectly. These phenomena eventually will degrade the capacity, cycle life and thermal stability of LiBs~\cite{woodford2013electrochemical,
bandhauer2011critical,pinson2013theory}. As such, too steep gradients should be circumvented during charging. This implies a necessity for restricting $V_s-V_b$. Besides, the restriction should be increasingly stricter as SoC grows, because a LiB becomes more vulnerable to a large Li-ion concentration gradient. The constraint about $V_s-V_b$ is then designed as an affine decreasing function of SoC:
\begin{align}\label{Vs-Vb-gamma12}
V_s-V_b \leq \gamma_1\rm{SoC}+\gamma_2,
\end{align}
where $\gamma_1\leq 0$ and $\gamma_2\geq 0$ are two coefficients. It can be rewritten as 
\begin{align}\label{inequality-eta}
\eta \leq \gamma_2,
\end{align}
where
\begin{align}\nonumber
\eta = -\frac{C_b+\gamma_1C_b+C_s}{C_b+C_s}V_b + \frac{C_b+C_s-\gamma_1C_s}{C_b+C_s}V_s.
\end{align}

\begin{remark}
The constraints in~\eqref{inequality-SoC}-\eqref{inequality-Vs} are either standard or can find equivalents in the literature. But~\eqref{Vs-Vb-gamma12} or~\eqref{inequality-eta} is unique as no similar ones have been considered in previous studies about MPC charging, despite their implications for enhancing the health consciousness in charging. Here, it is the NDC model that allows such a constraint to be applied. Furthermore, this model, as shown in~\cite{Tian:IECON:2018, tian2019NDC}, offers higher predictive accuracy than other popular ECMs, e.g., the Rint and Thevenin models, thus in a better position to ensure accurate charging control. These factors point to the advantage and appeal of using the NDC model to perform charging control design. 

\end{remark}



\section{Health-Aware Battery Charging via Explicit Model Predictive Control}\label{Sec:Charging-Strategy}

This section states an MPC-based charging control problem and then develops an eMPC-based charging control law. The latter effort will involve model linearization and computation of explicit solutions to the stated MPC problem.

\subsection{Health-Aware Charging Problem Formulation}\label{Sec:Health-aware-charging}

To begin with, let us convert the original state-space model~\eqref{state-space-equation} into a form that admits standard MPC formulation, which includes no direct input-output feedthrough in the measurement equation. To this end, define
\begin{align*} 
x &=\begin{bmatrix}V_b&V_s&I\end{bmatrix}^{\T},\\
y &= \begin{bmatrix}
\rm{SoC}& V_s & I & V & \eta
\end{bmatrix}^{\T},
\end{align*}
where $x$ is the state vector, and $y$ the output vector. Then, one can transform~\eqref{state-space-equation} by some manipulation and discretization into the following form:
\begin{subequations}\label{new-state-space-equation}
\begin{align}[left = \empheqlbrace\,] \label{new-state-equation}
x_{k+1}& = \mathcal{A}x_{k}+\mathcal{B}u_k,
\\ \label{new-output-equation}
y_k & = g(x_k), 
\end{align}
\end{subequations}
where $u_k=I_{k+1}-I_k$, 
\begin{align*}
\mathcal{A}=
\begin{bmatrix}
\tilde{A} & \tilde{B} \\
0 & 1
\end{bmatrix}, \ 
\mathcal{B}=
\begin{bmatrix}
0\\ 1 
\end{bmatrix},
\end{align*}
and $g(\cdot)$ represents the nonlinear mapping from $x$ to $y$. In above, $\tilde A$ and $\tilde B$ are the discretization-based counterparts of $A$ and $B$ under sampling interval $\Delta t$. 
Furthermore, the constraints in~\eqref{inequality-SoC}-\eqref{inequality-eta} can be put together in a compact form:
\begin{align}\nonumber
y_{\rm{min}} \leq y \leq y_{\rm{max}}.
\end{align}
In this setting, the health-aware charging control problem can be achieved by solving the following nonlinear MPC (NMPC) problem at time step $k$:
\begin{align}\label{Nonlinear-MPC-problem}
\min_{z} & \sum^{N-1}_{k=0} {1 \over 2}({\rm SoC}_k-{\breve r})^{\T}Q({\rm SoC}_k-{\breve r})+
{1 \over 2}\Delta u_k^{\T}R\Delta u_k, \\ \nonumber
\textrm {s.t.}&~\eqref{new-state-space-equation},\ x_0 = \breve{x}, \ u_{-1} = \breve{u},\\ \nonumber
& \ u_k = u_{k-1}+\Delta u_k, k = 0,\dots,N-1,\\ \nonumber
& \ \Delta u_k = 0, k=N_u,\dots,N-1,\\ \nonumber
& \ y_{\rm min}\leq y_k \leq y_{\rm max}, k = 0,\dots,N_c-1,
\end{align}
where $z=\begin{bmatrix} \Delta u_0& \dots &\Delta u_{N_u-1} \end{bmatrix}^{\T} \in \mathbb{R}^{N_u}$ is the future input sequence to be optimized, $\breve r$ the target SoC, $\breve{x}$ the model state at current time instant, and $\breve{u}$ the control input applied in the previous sampling interval, respectively. Besides, $N$ represents the prediction horizon, $N_u$ the input horizon, $N_c$ the constraint horizon, $Q=Q^{\T}\succeq 0$ and $R=R^{\T}\succ0$. The problem~\eqref{Nonlinear-MPC-problem} can be solved using nonlinear programming at each time instant. When the optimal solution $z^*$ is found, i.e.,
\begin{align} \label{optimizer-z}
z^*=\begin{bmatrix} \Delta u_0^*& \dots &\Delta u_{N_u-1}^* \end{bmatrix}^{\T},
\end{align}
its first element $\Delta u_0^*$ is used to compute
\begin{align}\nonumber
u_0 = \breve{u}+\Delta u_0^*.
\end{align}
The current to be applied for charging then is given by
\begin{align}\label{PWA-I}
I_1=u_0+I_0=\breve{u}+\Delta u_0^*+
\begin{bmatrix}0&0&1\end{bmatrix}
\breve{x}.
\end{align}
After this, the entire optimization problem is resolved at the next time instant with a new starting point.


The problem in~\eqref{Nonlinear-MPC-problem} gives a complete description of MPC-based charging control based on the NDC model. However, 
the online computation for the nonlinear programming is relatively formidable, which limits its applicability to embedded charging control. Hence, it is our aim to address~\eqref{Nonlinear-MPC-problem} {\em via} eMPC for easy online computation.  As eMPC is designed for linear systems, one must first linearize~\eqref{new-state-space-equation}, and the linearization is concerned with $h(V_s)$ and $R_0(V_s)$. Note that a single linear function is not accurate enough to approximate it as $V_s$ changes in charging. This motivates us to adopt multi-segment linear approximation to enhance the approximation accuracy. 

Proceeding to show this idea, consider linearizing $h(V_s)$ and $R_0(V_s)$ around a general fixed operating point, $V_s^{\rm{op}}$, as a first step. For $h(V_s)$, a linear approximation is considered, i.e.,
\begin{align}\label{linear-hVs}
h(V_s) \approx \lambda_1V_s+\lambda_2,
\end{align}
where 
\begin{align*}
\lambda_1=\left.\frac{\partial h(V_s)}{\partial V_s}\right\vert_{V_s=V_s^{\rm{op}}}, \ \lambda_2=h(V_s^{\rm{op}})-\lambda_1V_s^{\rm{op}}.
\end{align*}
For $R_0(V_s)$, it is approximated as a constant, i.e.,
\begin{align}\label{linearized-R0}
R_0(V_s)\approx R_0(V_s^{\mathrm{op}}).
\end{align}
By~\eqref{linear-hVs}-\eqref{linearized-R0}, one can modify~\eqref{new-output-equation} into a linear form as:
\begin{align}\label{linear-output-equation}
y_k = \mathcal{C}x_k+\mathcal{D},
\end{align}
where
\begin{align*}
\mathcal{C} = \begin{bmatrix}
\frac{C_b}{C_b+C_s}&\frac{C_s}{C_b+C_s}&0\\
0&1&0\\
0&0&1 \\
0&\lambda_1&R_0(V_s^{\mathrm{op}})\\
-\frac{C_b+\gamma_1C_b+C_s}{C_b+C_s}&\frac{C_b+C_s-\gamma_1C_s}{C_b+C_s}&0
\end{bmatrix},\
\mathcal{D} = \begin{bmatrix}
0\\
0\\
0 \\
\lambda_2\\
0
\end{bmatrix}.
\end{align*}
Accordingly, the original nonlinear MPC problem~\eqref{Nonlinear-MPC-problem} would reduce to a linear one, which can be expressed as
\begin{align}\label{Linear-MPC-problem}
\min_{z} & \sum^{N-1}_{k=0} {1 \over 2}({\rm SoC}_k-{\breve r})^{\T}Q({\rm SoC}_k-{\breve r})+
{1 \over 2}\Delta u_k^{\T}R\Delta u_k, \\ \nonumber
\textrm{s.t.}&~\eqref{new-state-equation},~\eqref{linear-output-equation},\ x_0 = \breve{x}, \ u_{-1} = \breve{u},\\ \nonumber
& \ u_k = u_{k-1}+\Delta u_k, k = 0,\dots,N-1,\\ \nonumber
& \ \Delta u_k = 0, k=N_u,\dots,N-1,\\ \nonumber
& \ y_{\rm min}\leq y_k \leq y_{\rm max}, k = 0,\dots,N_c-1.
\end{align}
Next is to extend this procedure to multi-segment approximation. Specifically, one can select multiple linearization points, denoting them as $V_{s,i}^{\rm{op}}$ for $i=1,2,\cdots,N_{\rm{op}}$. The range of $V_s$ then is subdivided into $N_{\rm{op}}$ partitions. The same procedure as in~\eqref{linear-hVs}-\eqref{linear-output-equation} can be repeated for each $V_{s,i}^{\rm{op}}$. Finally, a set of linear MPC subproblems akin to~\eqref{Linear-MPC-problem} will be obtained.


\begin{figure*}[t]
\centering
\includegraphics[trim = {0 0 0 0}, clip, width=0.83\textwidth]{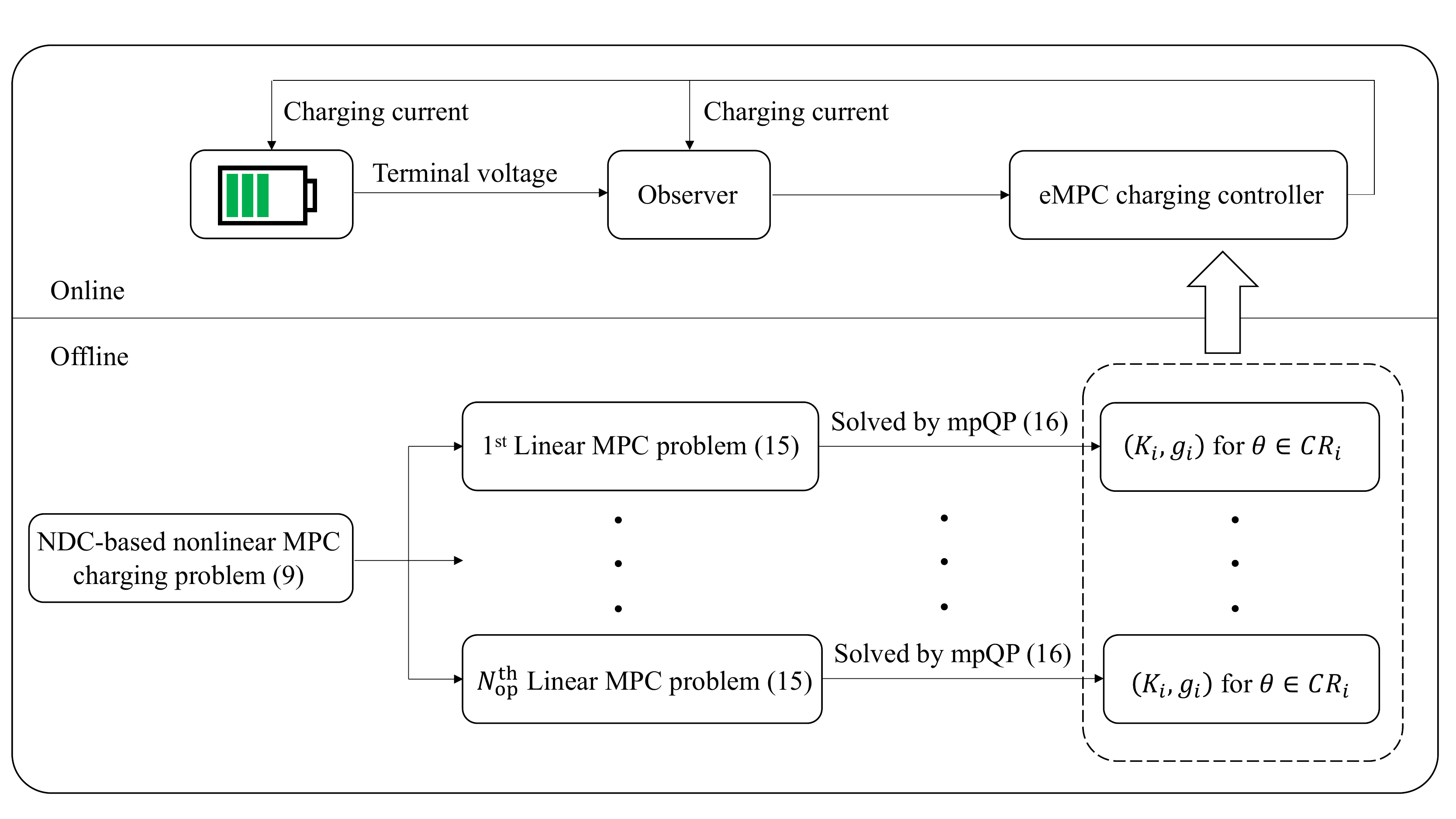}
\caption{Development of the eMPC-based charging control algorithm.} 
\label{Fig:Algorithm-plot}
\vspace{-4mm}
\end{figure*}

\begin{remark}\label{Constraint-Satisfaction}
The above procedure decomposes the original nonlinear MPC problem into a set of linear MPC problems with each based on a locally linearized model. In general, it is possible that constraint violation may happen upon model switching during the execution of multiple linear MPCs. But this issue does not cause much concern here. First, the linearization can be very precise. For LiB cells, the function $h(\cdot)$, which characterizes the SOC-OCV curve, is roughly composed of several almost flat regions, thus well lending itself to multi-segment linearization. The accuracy can be also easily improved further by using more operating points. Second, one can use the upper end of the segment as $V_s^{\rm op}$ to evaluate and approximate $R_0(V_s)$. Then $R_0(V_s)$ is replaced by a number larger than it should be. Such a conservatism will effectively reduce the chance of constraint violation. 

\end{remark}

\subsection{Charging Control Based on eMPC}\label{Sec:Explicit-MPC}


Consider the linear MPC charging control problem~\eqref{Linear-MPC-problem}. Based on~\eqref{new-state-equation} and~\eqref{linear-output-equation}, one can obtain
\begin{align*}
y_k & = \mathcal{C}\mathcal{A}^k\breve{x}+\sum^{k-1}_{j=0}\mathcal{C}\mathcal{A}^j\mathcal{B}u_{k-1-j}+\mathcal{D},\\
u_k & = \breve{u}+\sum^{k-1}_{j=0}\Delta u_j.
\end{align*}
Further, define the following vector of parameters:
\begin{align}\nonumber
\theta = \begin{bmatrix} \breve x^{\T} & \breve r^{\T} & \breve u^{\T} \end{bmatrix}^{\T}\in \mathbb{R}^{m}.
\end{align}
The optimization problem~\eqref{Linear-MPC-problem} then can be recast as a convex quadratic program (QP) taking a standard form~\cite{bemporad2015multiparametric}:
\begin{subequations}\label{mp-QP-problem}
\begin{align}\label{mp-QP-cost-function}
\min_{z} & \ {1 \over 2}z^{\T}\Sigma z+\left(F\theta \right)^{\T}z, \\ \label{mp-QP-constraint}
\textrm {s.t.}& \ Gz\leq S\theta+W,
\end{align}
\end{subequations}
where $z\in \mathbb{R}^{N_u}$, $\Sigma\succ0\in \mathbb{R}^{N_u\times N_u}$ and $F \in \mathbb{R}^{N_u\times m}$. 

The QP problem~\eqref{mp-QP-problem} is also a multiparametric QP (mpQP) problem, as the characterization of its solution fundamentally involves the parameter vector $\theta$. The solution can be described as a set-valued function $Z^*(\theta): \Theta \rightarrow 2^{\mathbb{R}^n}$, where $Z^*(\theta)$ is a set of optimizer functions $z^*(\theta)$, $\Theta$ the set of feasible parameters (parameters that allow a non-empty set of $z$ to satisfy~\eqref{mp-QP-constraint}), and $2^{\mathbb{R}^n}$ the set of subsets of $\mathbb{R}^n$. 
It has been proven in~\cite{bemporad2002explicit} that, $\Theta$, which is provably a polyhedral set, can be partitioned into convex polyhedral regions, also referred to as critical regions and denoted as $CR_i$ for $i=1,2,\cdots,N_{CR}$. For each critical region,
\begin{align*}
z^*(\theta) = K_i \theta+ g_i, \ \forall \theta \in CR_i.
\end{align*}
In other words, $Z^*(\theta)$ is PWA and continuous over $\Theta$. An immediate implication is that, given the formulated problem, the charging control law will be PWA functions of the present charging state, the target SoC and the input in last time instant.

A few mpQP algorithms have been developed in the literature. Such an algorithm usually has a two-fold functionality: determining the partition of $\Theta$ into critical regions $CR_i$, and finding out the control law $z^*(\theta)$ associated with each $CR_i$. To design them, an important approach is the so-called geometric approach~\cite{Borrelli:CUP:2017}. It 1) decides a critical region around a specified parameter by using the sufficient and necessary conditions for optimality, 2) solves the mpQP for this region, and 3) partitions the rest of the feasible parameter space and continues the optimization until the space is fully explored. The literature also contains some other approaches, and an interested reader is referred to~\cite{Borrelli:CUP:2017} for a review. 

\begin{table*}[t]\centering
\begin{threeparttable}
\caption{Battery model parameters.}
\begin{tabular}{cccccccccccccc}
\toprule%
Name& $C_b/F$ &$C_s/F$ &$R_b/\Omega$&$R_s/\Omega$ &$\alpha_0$& $\alpha_1$ 
&$\alpha_2$ &$\alpha_3$ & $\alpha_4$ &$\alpha_5$ &$\beta_1$&$\beta_2$&$\beta_3$ \\
\midrule
Value & 9,913& 887& 0.025 & 0&3.2&3.041& -11.475 & 24.457& -23.536&8.513&0.09&0.35&10\\
\bottomrule
\end{tabular}
\label{Tab:Model-Parameters}
\end{threeparttable}
\vspace{-2mm}
\end{table*}

Putting together the above developments, the eMPC-based charging control algorithm is summarized as follows:

\par\noindent\rule{0.49\textwidth}{1pt}

\begin{itemize}
\item {\em Offline mpQP computation}
\begin{itemize}
\item Consider the first linear model
\begin{itemize}
\item Select a parameter $\theta_0$
\item Determine the critical region in the neighborhood of $\theta_0$, and denote it as $CR_0$
\item Solve the mpQP problem~\eqref{mp-QP-problem} to obtain $z^*(\theta) = K_0 \theta + g_0$ for $\theta \in CR_0$
\item Partition the parameter space outside $CR_0$, and determine $z^*(\theta)$ for new critical regions

\item Repeat the procedure until when the entire parameter space has been explored

\item Store in a table all $(K_i, g_i)$ for $i=1,2,\cdots,N_{CR}$
\end{itemize}

\item Repeat the procedure for all the other linear models

\end{itemize}

\item {\em Online eMPC-controlled charging}
\begin{itemize}

\item Determine the governing linear model at every time step $k$

\item Given $\theta_k$, search the stored $(K,g)$ table to find $CR_j$ such that $\theta_k \in CR_j$

\item Determine $z^*(\theta) = K_j \theta_k +g_j$ as well as the charging current $I_{k+1}$ as in~\eqref{PWA-I}

\item Repeat the charging control procedure until the condition for charging completion is satisfied
\end{itemize}
\end{itemize}
\par\noindent\rule{0.49\textwidth}{1pt}

The following remarks summarize further discussion of the above eMPC-based charging control algorithm.


\begin{remark}\label{Remark:Complexity}
The eMPC-based charging control algorithm moves the constrained optimization, which is computationally expensive, from online to offline by deriving explicit solutions to the considered optimization problem. The online control run at every time step involves only searching through a look-up table comprising critical regions to find out the correct PWA function and then evaluating it. Our analysis, together with instructions in~\cite{bemporad2018model}, shows that the computational and storage costs in the control run are generally affordable for embedded   systems. Further, the literature also contains studies about efficient eMPC implementation by minimizing the time to evaluate the PWA functions and   reducing the    memory needed to store numbers, e.g.,~\cite{tondel2003evaluation,gersnoviez2017high}. They can be integrated with our work to enable highly efficient   charging   in practice. Therefore, the eMPC-based charging control algorithm would yield substantial online computational economy and implementability on relatively low-end computing hardware,   filling a gap that exists in previous research on MPC-based charging. 
\end{remark}


\begin{remark}
The above design assumes state-feedback design for convenience of discussion. It can be easily extended to output-feedback control, which is necessary in practice as the internal states of the NDC model, $V_b$ and $V_s$, cannot be measured. To attain this end, one can just use a nonlinear state observer to perform real-time state estimation. Then, the charging control setup is a closed loop between the LiB, observer and eMPC controller, which is outlined in Figure~\ref{Fig:Algorithm-plot}. To further illustrate this, Section~\ref{Sec:EKF-eMPC} offers a case study that uses the extended Kalman filter (EKF) for state estimation and combines it with the proposed eMPC control law to perform output-feedback charging control. 
\end{remark}

\begin{remark}\label{State-Feedback}
The proposed design allows for extension to a more sophisticated model. For example, if a thermal model is coupled with the NDC model, one can follow the design approach to do linearization and then conduct eMPC design to enable temperature-conscious charging control. Such a treatment can also be modified to deal with the case when the model parameters are temperature-dependent. It is also noteworthy that the above can be applied to other ECMs, such as the Thevenin model, with custom-built eMPC charging control algorithms. 
\end{remark}

\begin{table}[t]\centering
\begin{threeparttable}
\caption{Linearization setting.}
\begin{tabular}{ccccccc}
\toprule%
No. &$V_s$& $V_s^{\rm{op}}$ &$\lambda_1$ &$\lambda_2$&$R_0$ &$N_{CR}$\\
\midrule
I&$[0.20,0.50]$& 0.39&0.6505 &3.3701& 0.091&15\\
II&$[0.50,0.60]$&0.60 &0.8659 & 3.2685& 0.096&14 \\
III&$[0.60,0.70]$&0.70 &0.8562 &3.2752& 0.107 &14 \\
IV&$[0.70,0.74]$&0.74 &0.8503 &3.2794& 0.116&14 \\
V&$[0.74,0.78]$&0.78 &0.8581 & 3.2734& 0.129&14 \\
VI&$[0.78,0.81]$&0.81 &0.8810 &3.2551& 0.142&15 \\
VII&$[0.81,0.84]$&0.84&0.9259 &3.2181 & 0.161&15 \\
VIII&$[0.84,0.87]$&0.87 &1.0002 &3.1544 & 0.185&15 \\
IX&$[0.87,0.90]$&0.90 &1.1123 & 3.0551 & 0.219&15 \\
\bottomrule
\end{tabular}
\label{Tab:Operating-Points}
\end{threeparttable}
\vspace{-2mm}
\end{table}

\begin{figure} [t]
\centering 
\subfigure[Piecewise linearization of $h(V_s)$]
{\includegraphics[trim = {0 0 0 0}, clip, width=0.4\textwidth]{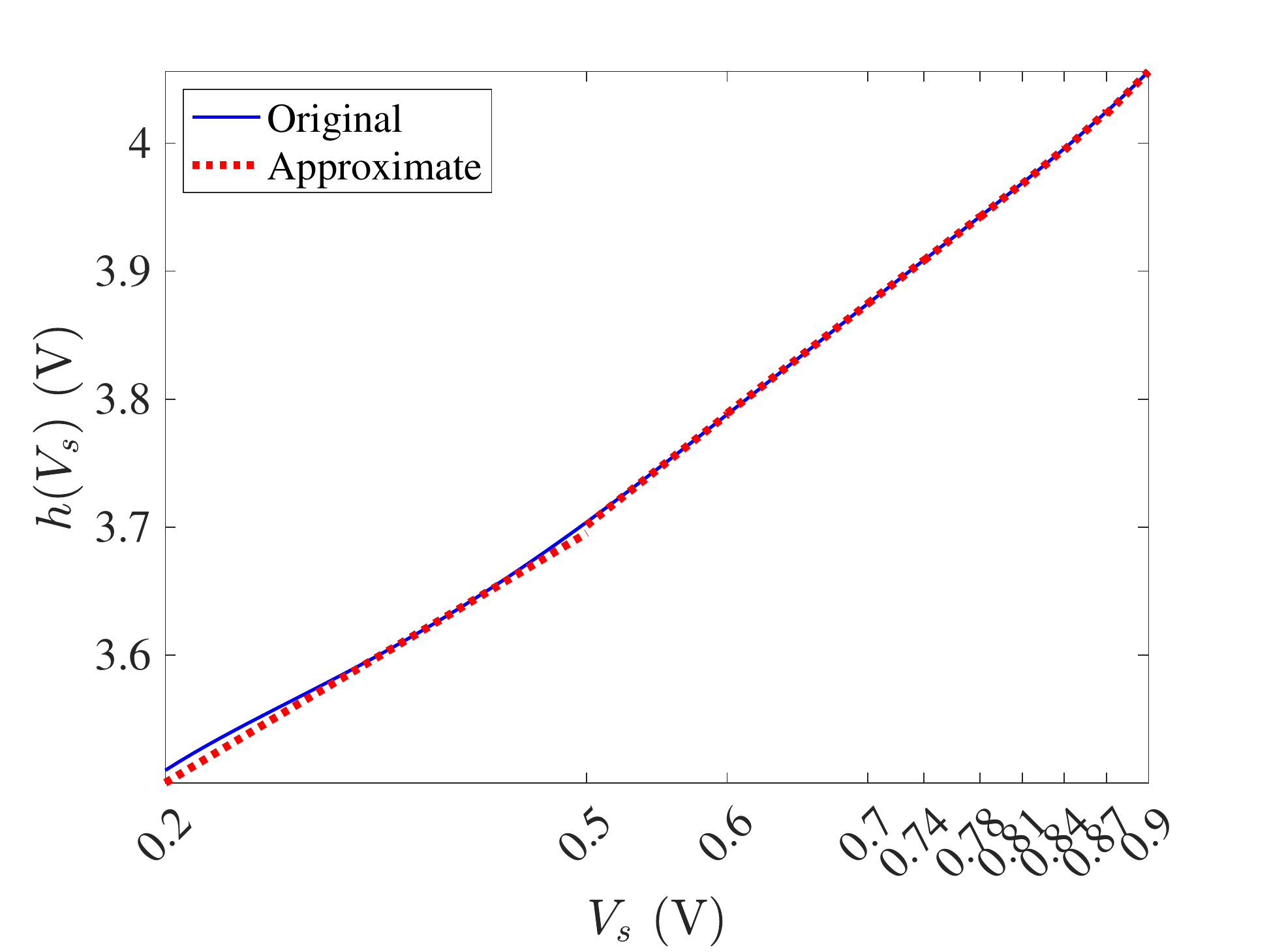}\label{Fig:linear-hVs}} \\
\vspace{-2mm}
\hspace{0in} 
\subfigure[Constant-wise approximation of $R_0(V_s)$]
{\includegraphics[trim = {0 0 0 0}, clip, width=0.4\textwidth]{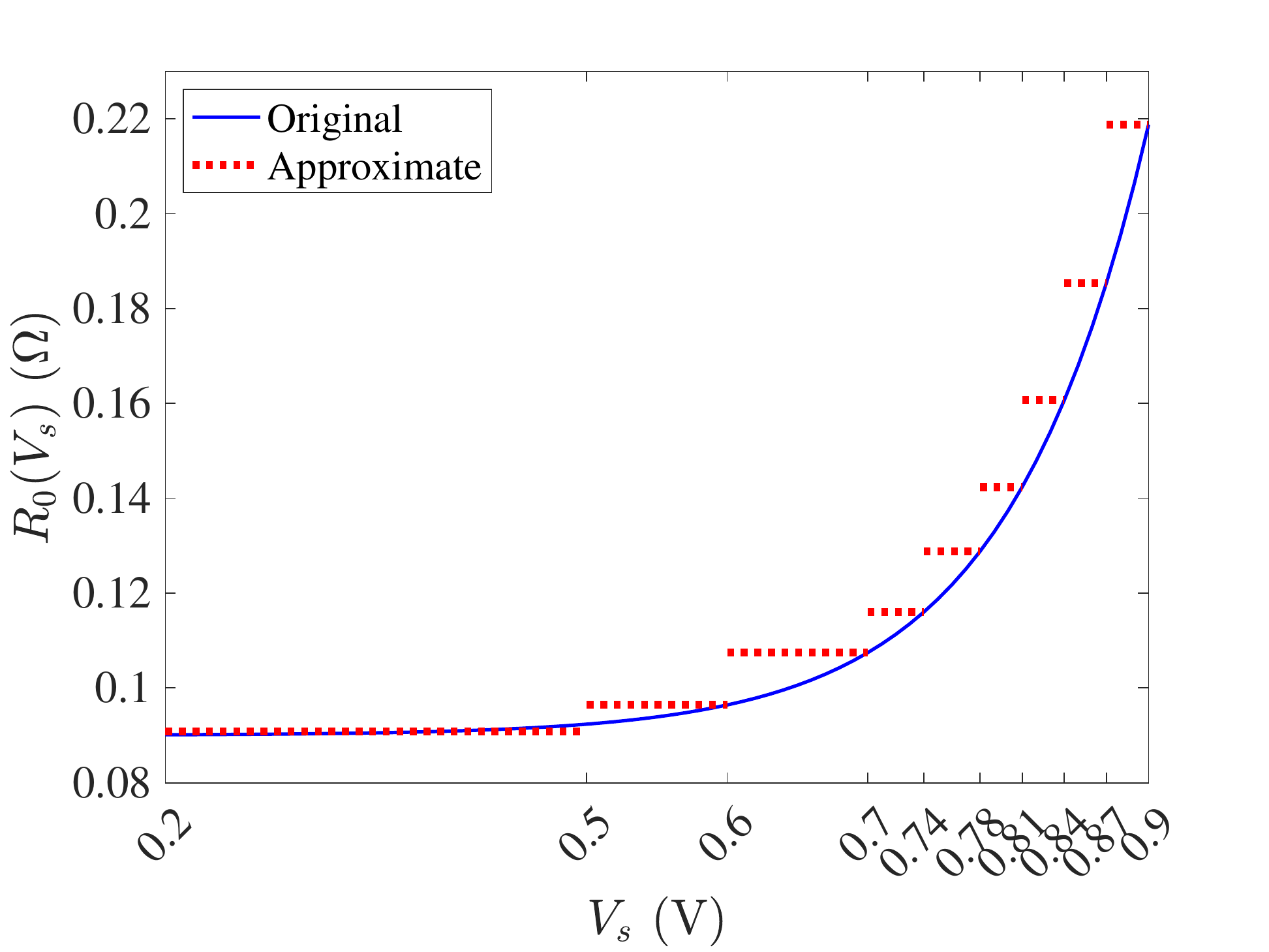}\label{Fig:linear-R0}}
\caption{Multi-segment approximation of $h(V_s)$ and $R_0(V_s)$.}
\label{Fig:Linearization-R0-hVs}
\vspace{-5mm}
\end{figure}

\begin{figure}[t]
\centering
\includegraphics[trim = {0 0 0 0}, clip, width=0.4\textwidth]{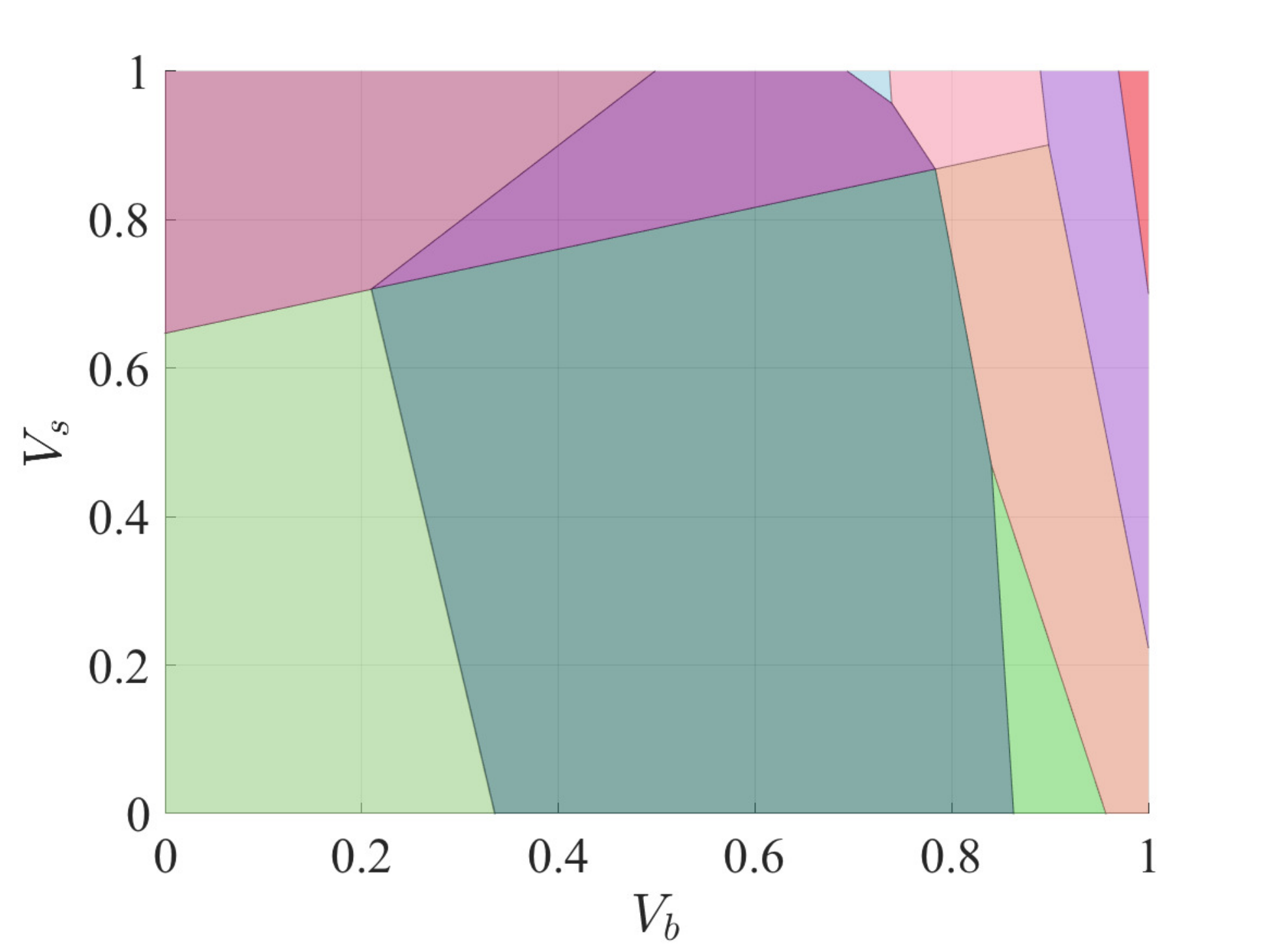}
\caption{Critical region partitioning on the $V_b$-$V_s$ plane when $I=2$~A, $\breve{r}=0.9$, and $\breve{u}=0$~A.}
\label{Fig:2D-plot}
\vspace{-7mm}
\end{figure}

\section{Numerical Simulation}\label{Sec:Numerical-Simulation}

This section presents simulation results to validate the proposed eMPC-based charging control algorithm. It offers an overview of the algorithm first through a basic case study and then investigates its performance in different settings.


\subsection{Basic Case Study}\label{Sec:Case-Study}
Given a 3 Ah LiB cell governed by the NDC model with the parameters shown in Table~\ref{Tab:Model-Parameters}, consider the optimal charging problem based on the formulation in~\eqref{Nonlinear-MPC-problem}. Here, the charging objective is to raise the SoC from $20\%$ to $90\%$ under the following constraints:
\begin{gather} \nonumber
V_s\leq 0.95~{\rm V}, \ 0~{\rm A}\leq I\leq 3~{\rm A},\ V\leq 4.2~{\rm V},\\ \nonumber
V_s-V_b \leq -0.04\cdot {\rm SoC}+0.08.
\end{gather}
The sampling interval $\Delta t$ for model discretization is 1 min. For the eMPC run, $Q=1$, $R = 0.1$, the prediction horizon $N=10$, the control horizon $N_u=2$, and the contraint horizon $N_c=2$ for~\eqref{Vs-Vb-gamma12} and $N_c=1$ for the other constraints.

\begin{figure*} [t]
\centering 
\subfigure[Charging current profile]
{\includegraphics[trim = {0 0 0 0}, clip, width=0.3\textwidth]{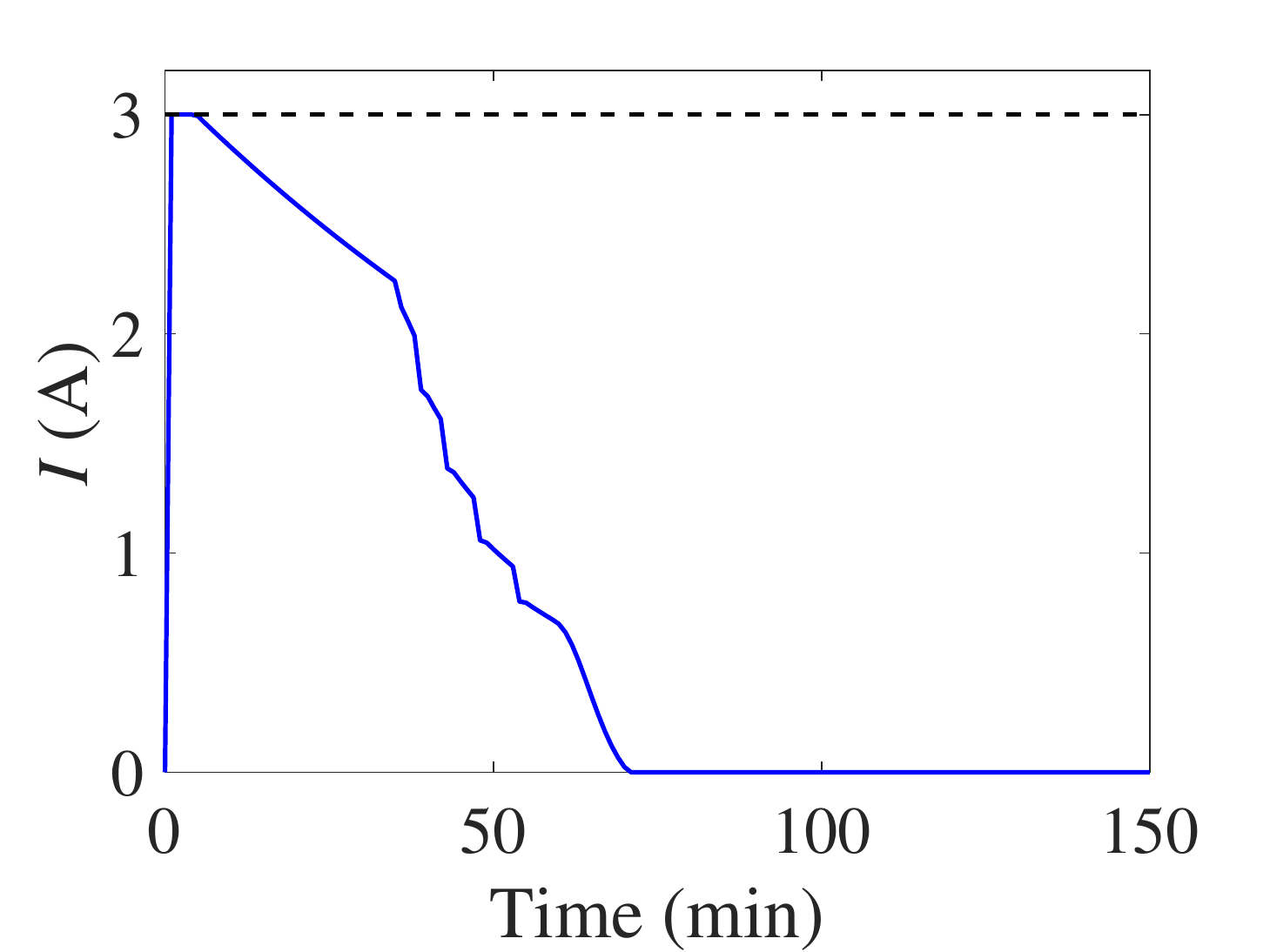}\label{Fig:current-simulation}} 
\hspace{0in} 
\subfigure[SoC profile]
{\includegraphics[trim = {0 0 0 0}, clip, width=0.3\textwidth]{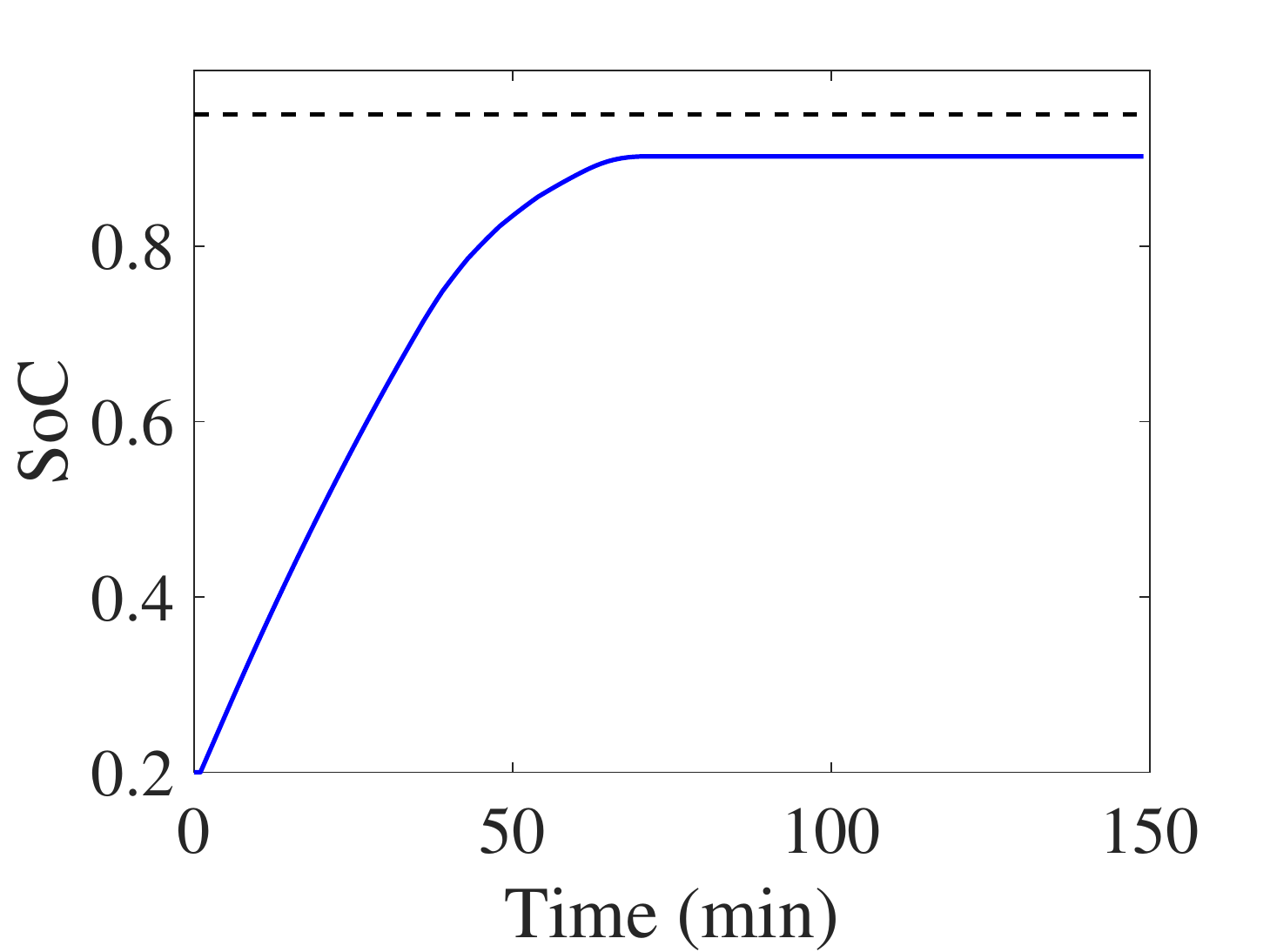}\label{Fig:SoC-simulation}}
\hspace{0in} 
\subfigure[Voltage profile]
{ \includegraphics[trim = {0 0 0 0}, clip, width=0.3\textwidth]{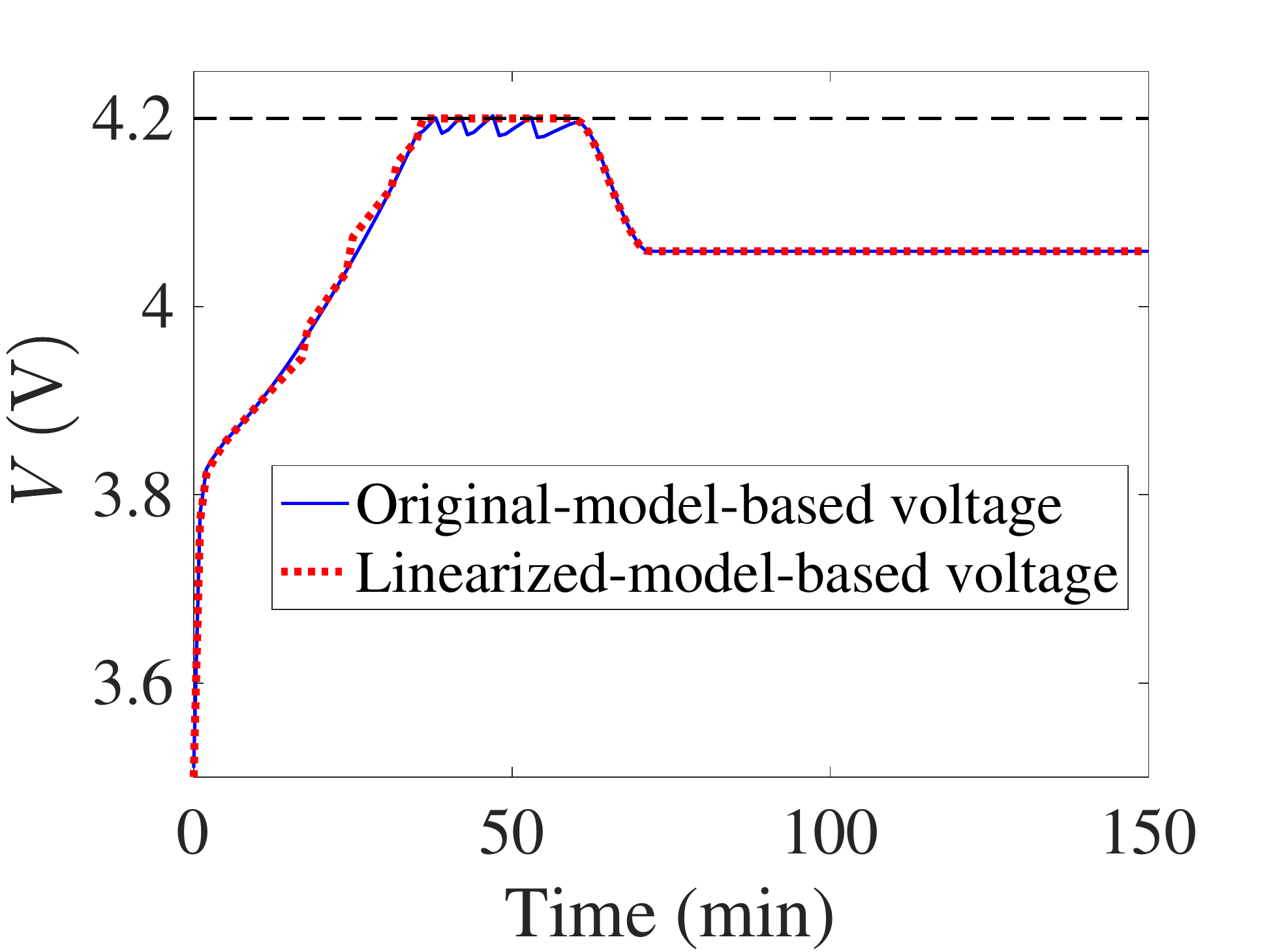}\label{Fig:voltage-simulation}} \\
\vspace{-3mm}
\subfigure[$V_s-V_b$]
{\includegraphics[trim = {0 0 0 0}, clip, width=0.3\textwidth]{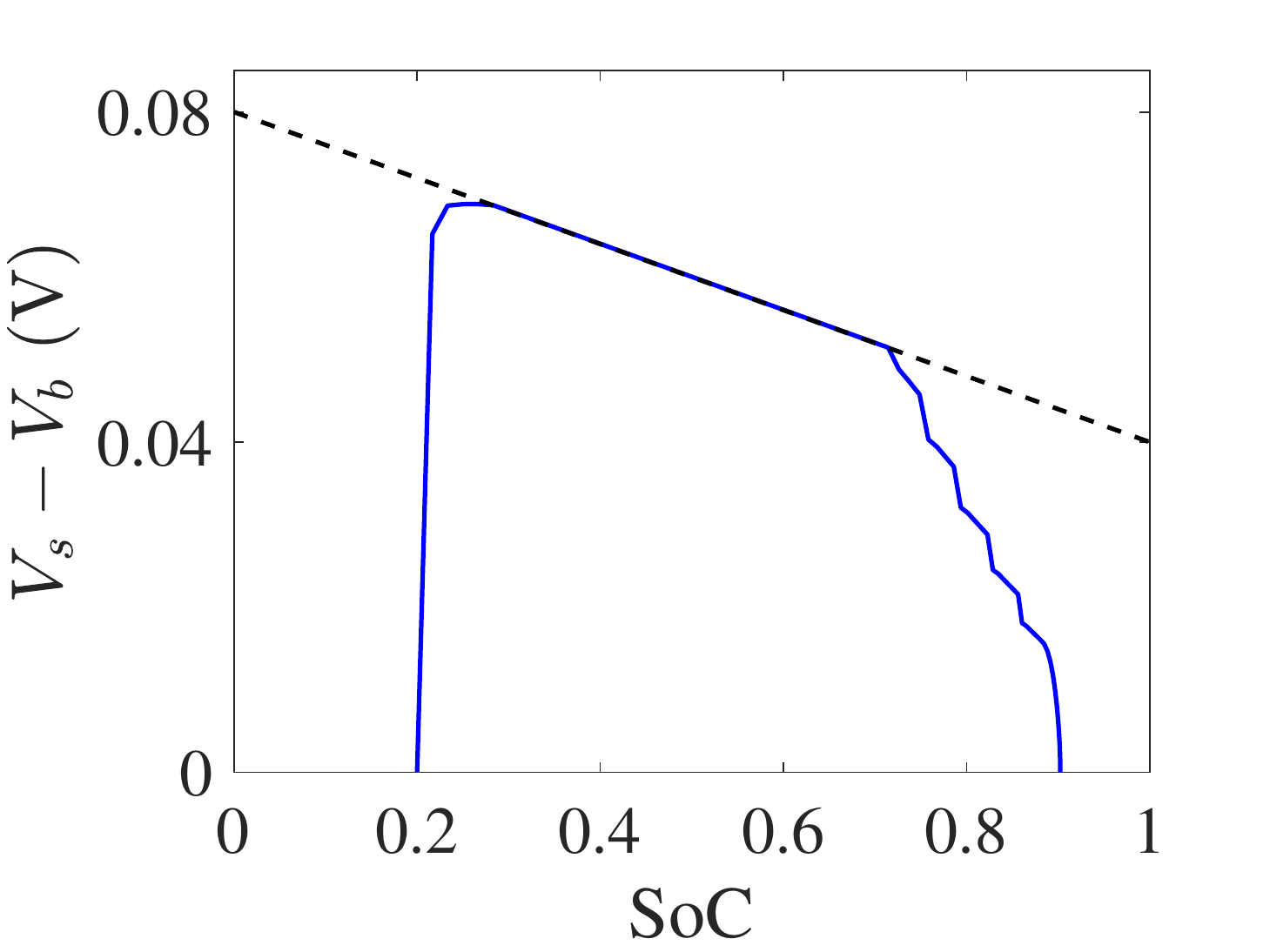}\label{Fig:Vs-Vb-simulation}}\hspace{0in} 
\subfigure[$V_b$ and $V_s$]
{\includegraphics[trim = {0 0 0 0}, clip, width=0.3\textwidth]{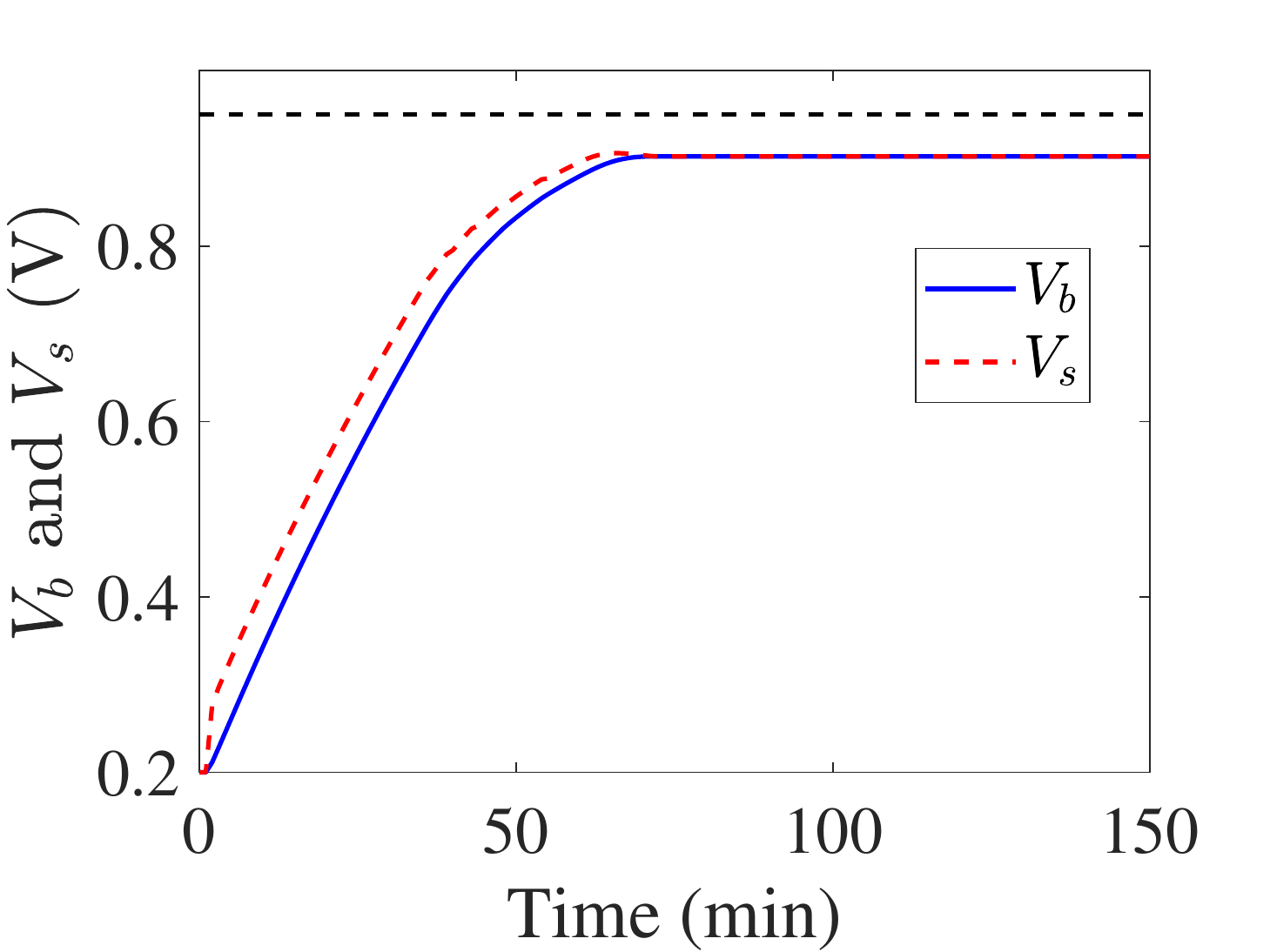}
\label{Fig:Vs-and-Vb}} 
\hspace{0in} 
\caption{A basic case study of the eMPC-based charging control (black dashed lines denote constraints).} 
\label{Fig:Charging-Results}
\vspace{-5mm}
\end{figure*}

\begin{figure*} [t]
\centering 
\subfigure[Charging current profile]
{\includegraphics[trim = {0 0 0 0}, clip, width=0.3\textwidth]{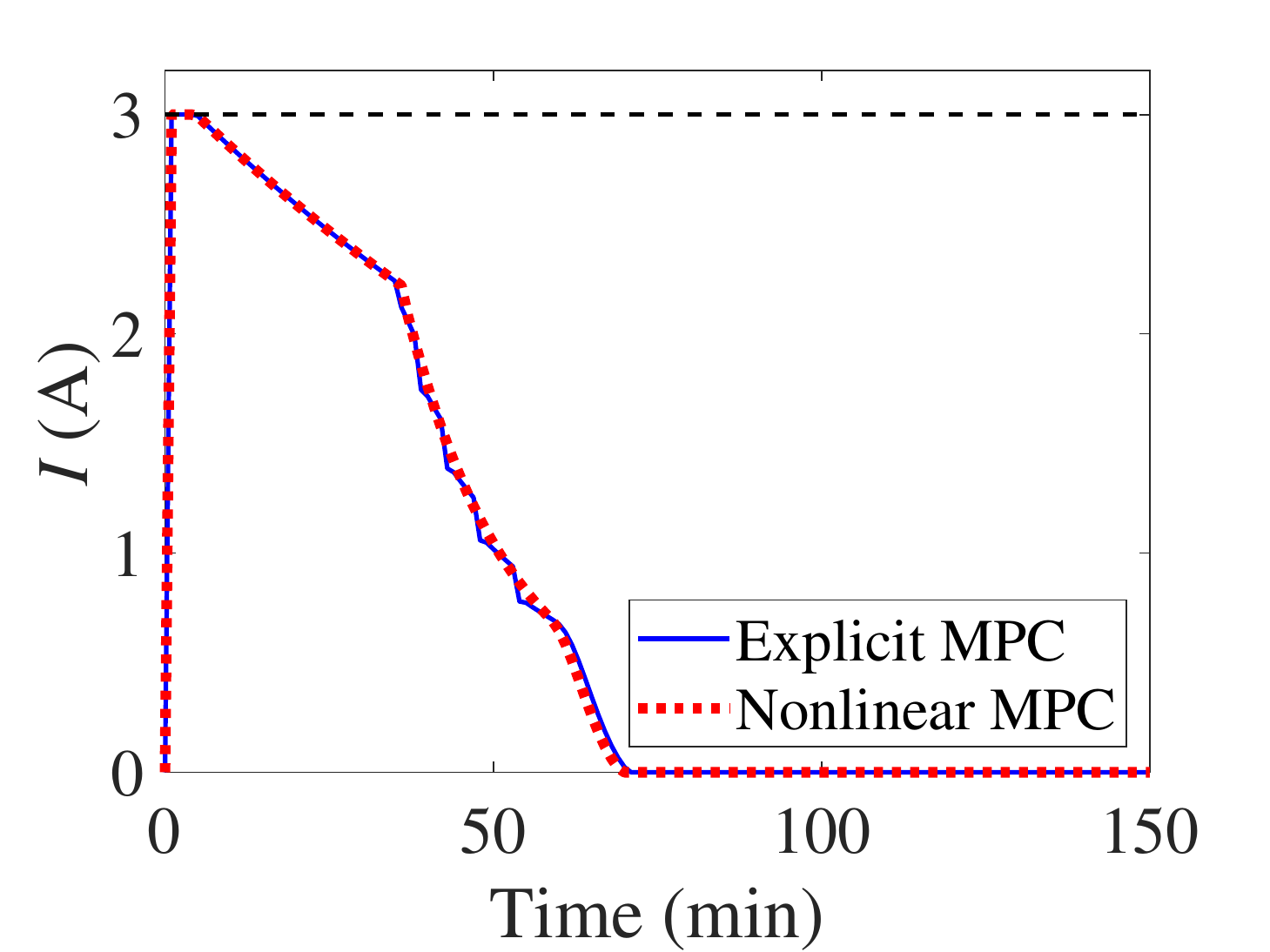}\label{Fig:current-ENMPC}} 
\hspace{0in} 
\subfigure[SoC profile]
{\includegraphics[trim = {0 0 0 0}, clip, width=0.3\textwidth]{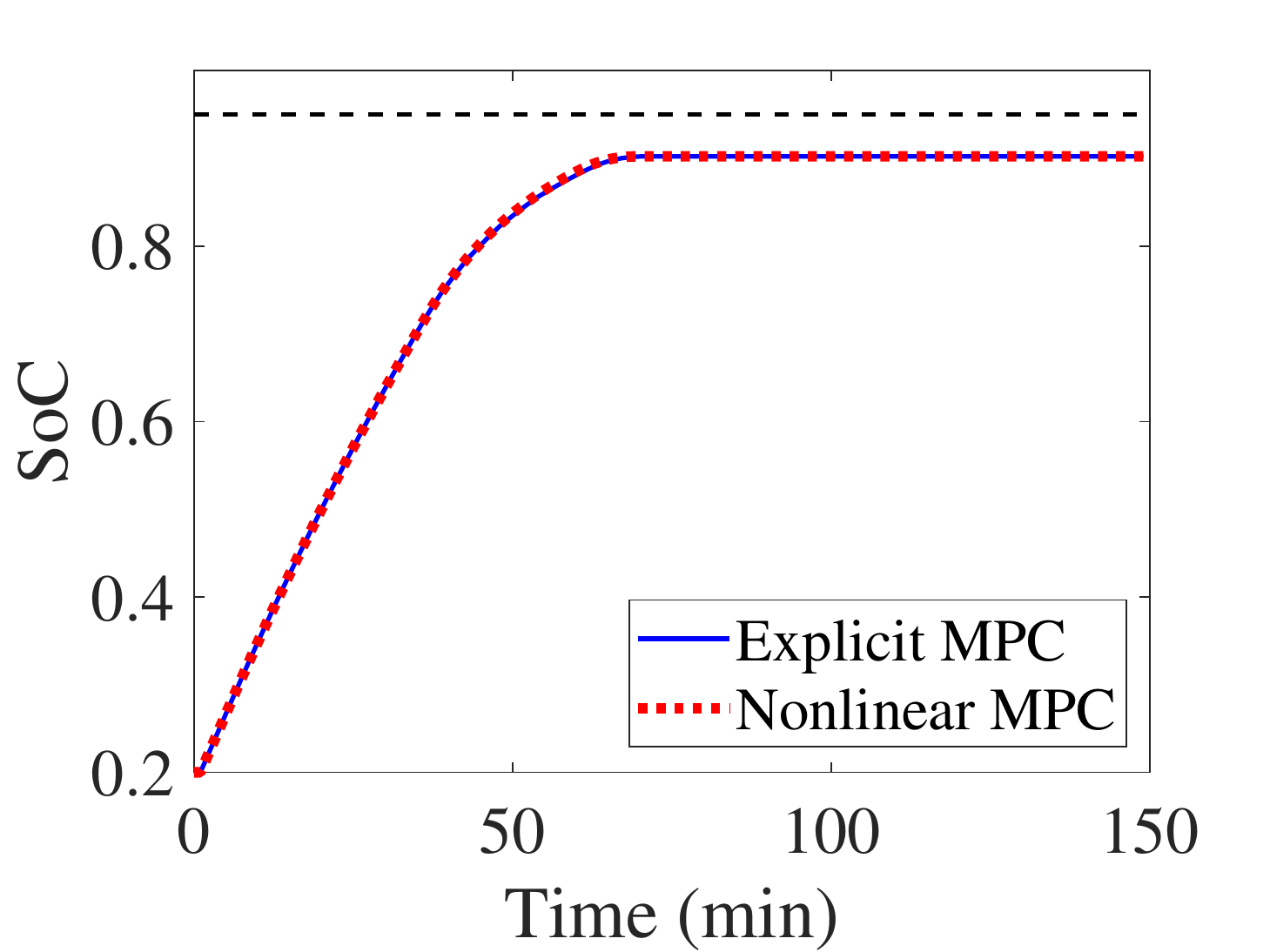}\label{Fig:SoC-ENMPC}}
\hspace{0in} 
\subfigure[Voltage profile]
{ \includegraphics[trim = {0 0 0 0}, clip, width=0.3\textwidth]{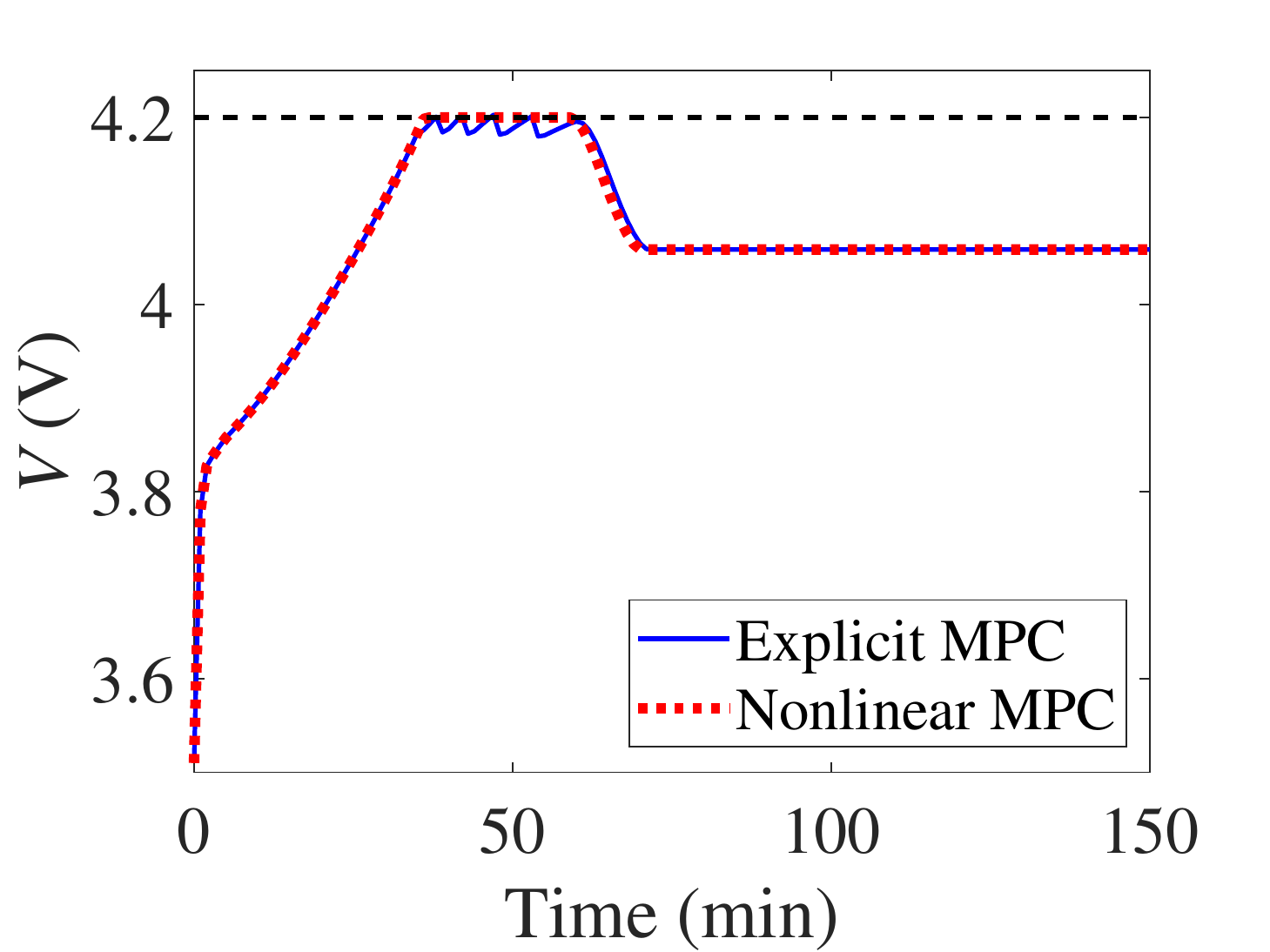}\label{Fig:voltage-ENMPC}} 
\caption{Comparison between eMPC and NMPC for charging control.} 
\label{Fig:EMPC-VS-NMPC}
\vspace{-5mm}
\end{figure*}

\begin{figure*} [t]
\centering 
\subfigure[Charging current profile]
{\includegraphics[trim = {0 0 0 0}, clip, width=0.3\textwidth]{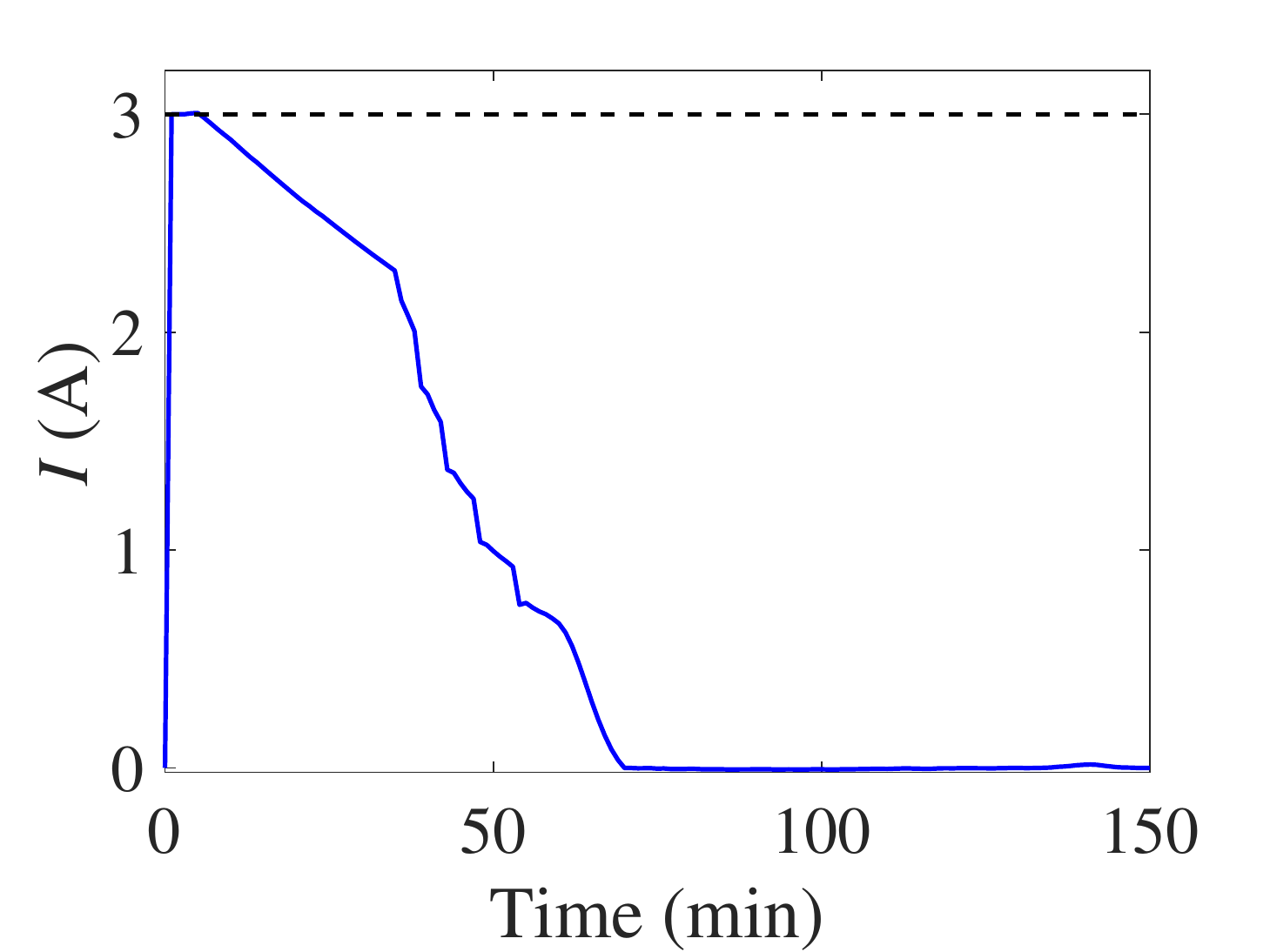}\label{Fig:EKF-current}} 
\hspace{0in} 
\subfigure[SoC profile]
{\includegraphics[trim = {0 0 0 0}, clip, width=0.3\textwidth]{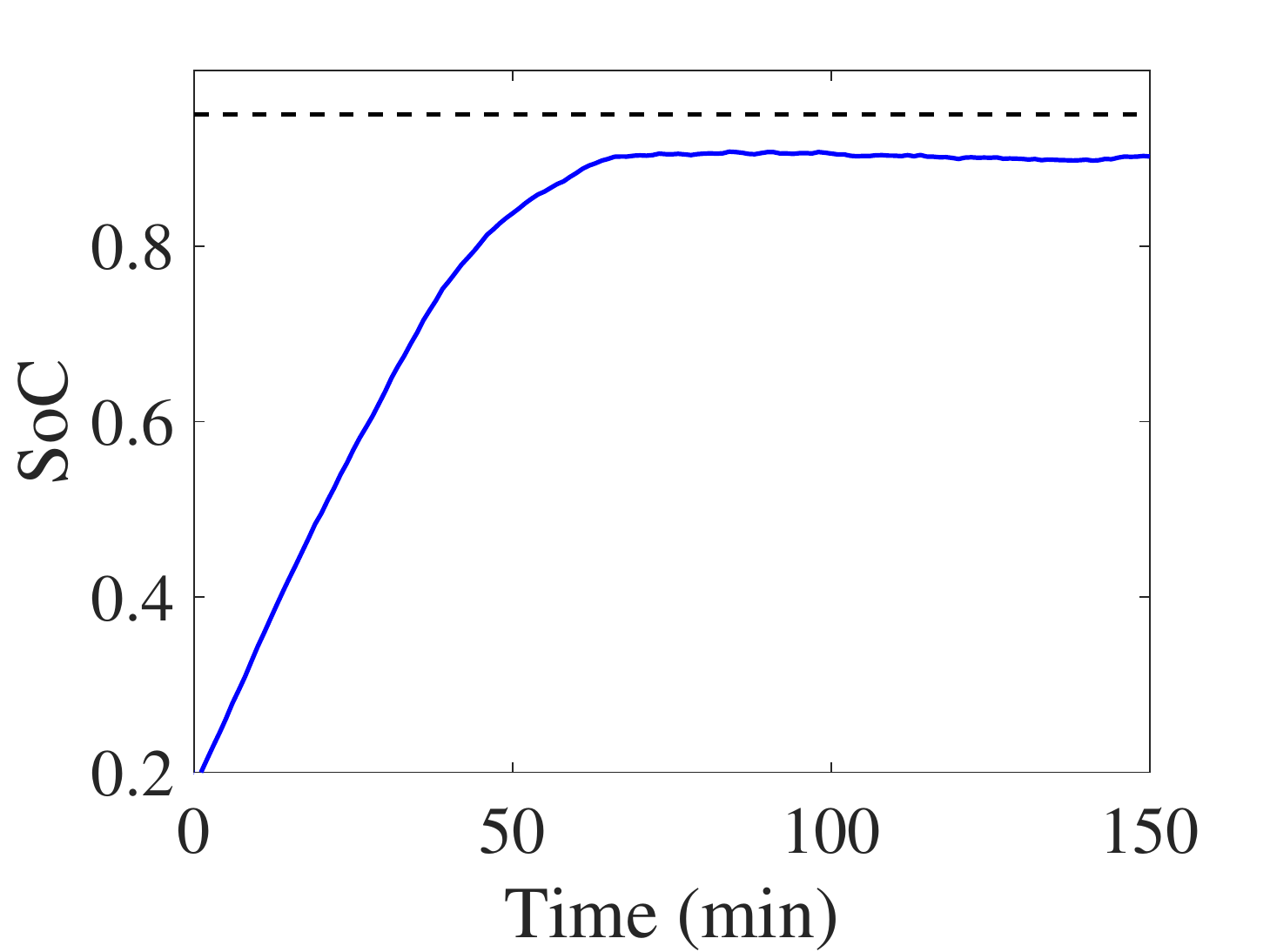}\label{Fig:EKF-SoC}}
\hspace{0in} 
\subfigure[Voltage profile]
{ \includegraphics[trim = {0 0 0 0}, clip, width=0.3\textwidth]{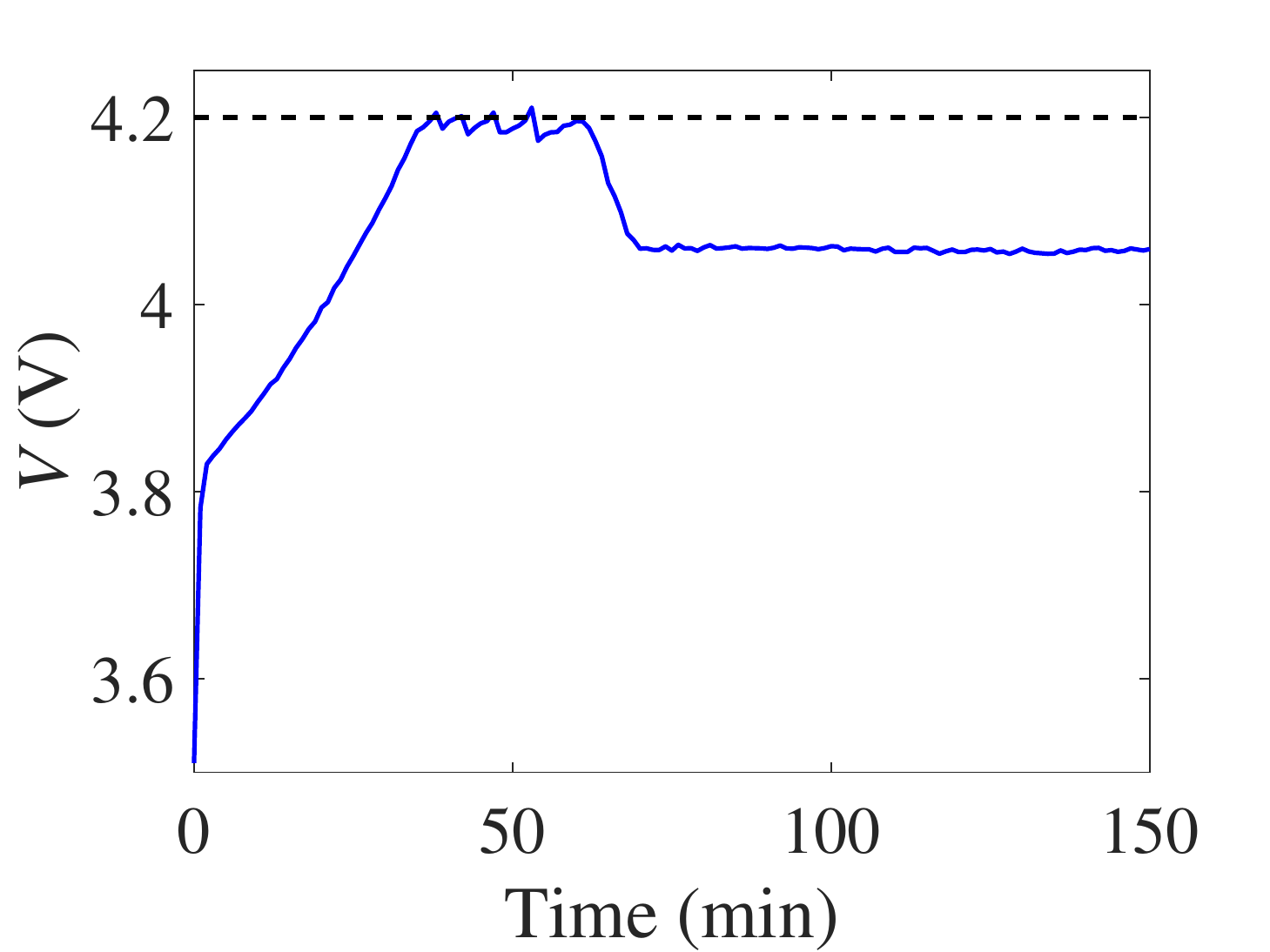}\label{Fig:EKF-voltage}} \\
\vspace{-3mm}
\subfigure[$V_s-V_b$]
{\includegraphics[trim = {0 0 0 0}, clip, width=0.3\textwidth]{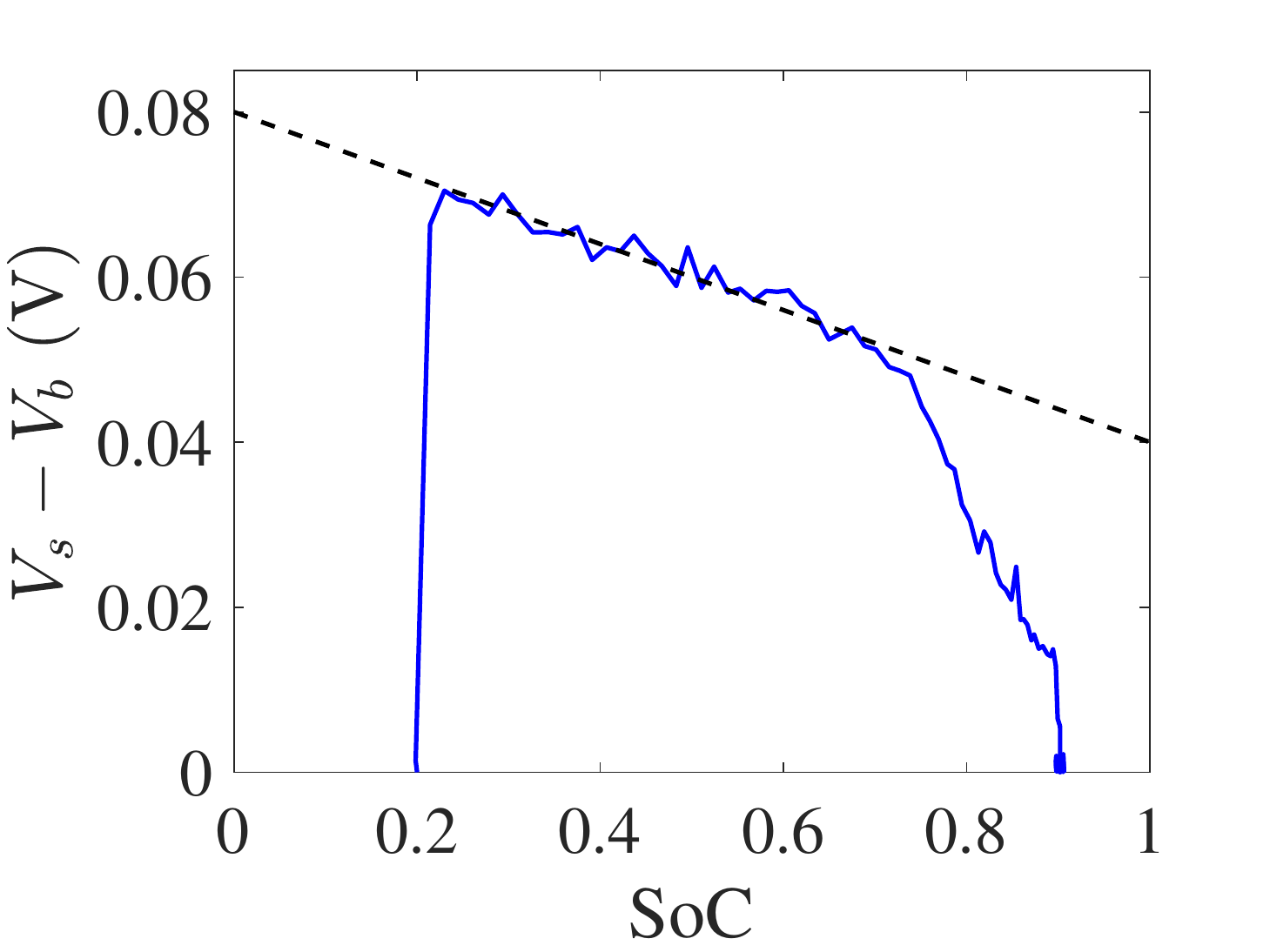}\label{Fig:EKF-Vs-Vb}}\hspace{0in} 
\subfigure[$V_b$ and $V_s$]
{\includegraphics[trim = {0 0 0 0}, clip, width=0.3\textwidth]{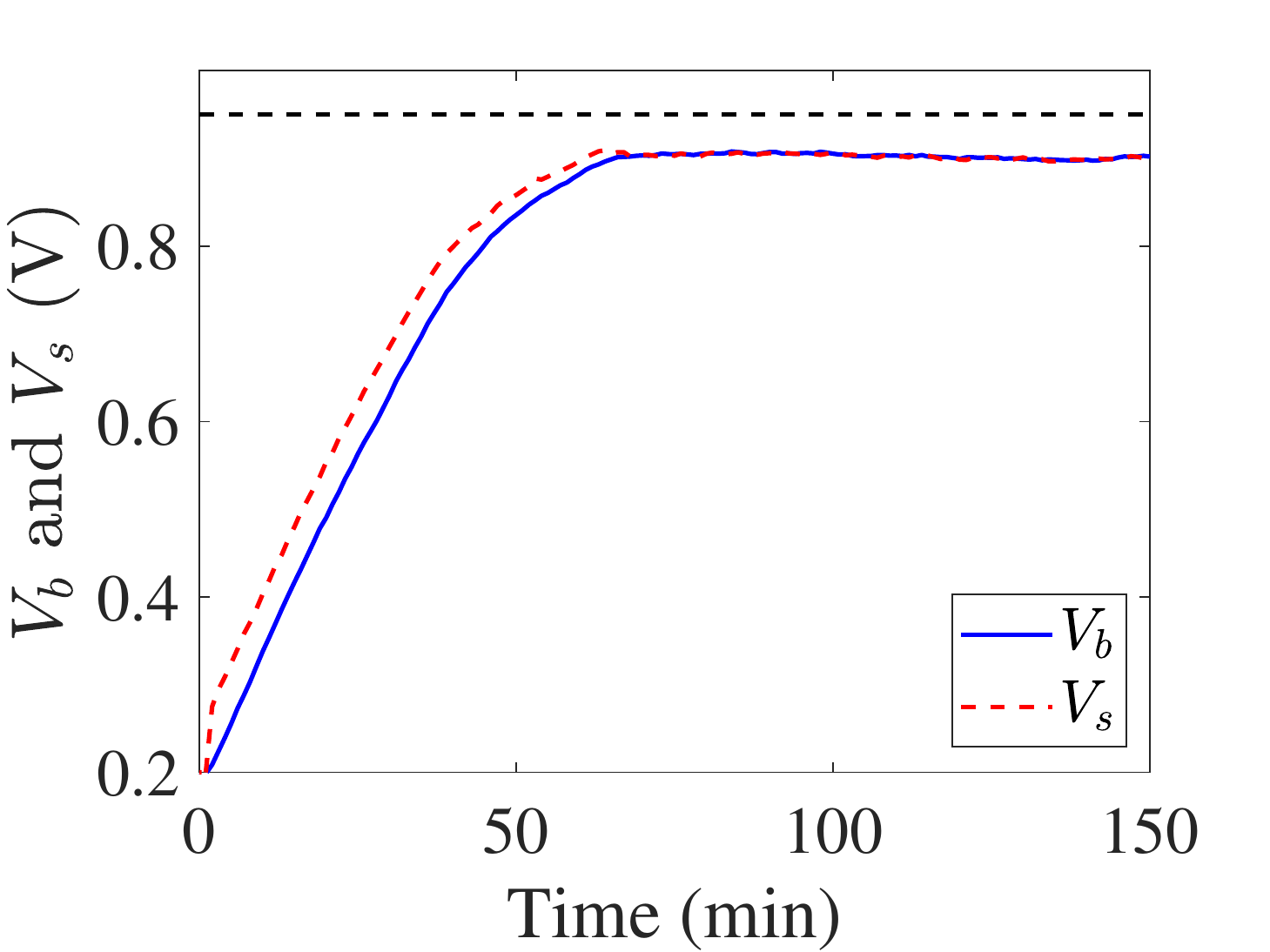}\label{Fig:EKF-Vs-and-Vb}} 
\hspace{0in} 
\caption{Output-feedback charging control based on eMPC and EKF.} 
\label{Fig:EKF-eMPC}
\vspace{-5mm}
\end{figure*}

Next, the nonlinear MPC problem~\eqref{Nonlinear-MPC-problem} is broken down into nine linear MPC problems through multi-segment approximation. Table~\ref{Tab:Operating-Points} summarizes the linearization setting, which includes the range of $V_s$ for each segment, the linearization point of $V_s$, and the obtained linearization results for $h(V_s)$ and $R_0(V_s)$. The approximation of $h(V_s)$ and $R_0(V_s)$ is depicted in Figure~\ref{Fig:Linearization-R0-hVs}. Thanks to the linearization, each linear MPC problem can be characterized in the form of~\eqref{Linear-MPC-problem}. One can then compute the explicit solution to every problem by conveniently resorting to the MATLAB\textsuperscript{\textregistered} MPC Toolbox~\cite{bemporad2018model}, which leads to nine eMPCs that combine to make up the charging control algorithm. Note that the eMPCs have different numbers of critical region partitions, as they are based on different models and operating ranges. For each eMPC, the $N_{CR}$ is shown in Table~\ref{Tab:Operating-Points}. 
To give the reader a flavor of the control law, let us look at the second eMPC. For this one, the five-dimensional parameter space $\Theta$ is divided into 14 convex polyhedral critical regions, with each one associated with a PWA function of $\theta$. For example, the tenth region is given by
\begin{align*}\footnotesize
\begin{bmatrix}
0.25&-0.90&-0.35&0&0\\
0.40&0.42&0.02&-0.81&-0.06 \\
0.89&0.44&0.15&0&0 \\
-0.69&0.67&0.28&0&0 \\
0&1&0&0&0 \\
0&0&0&1&0 \\
0&0&0&0&1 \\
0&-1&0&0&0 \\
0&0&0&-1&0 \\
0&0&0&0&-1
\end{bmatrix}
\begin{bmatrix}
V_{b,k}\\V_{s,k}\\I_{k}\\\breve{r}\\u_{k-1}
\end{bmatrix}
\leq
\begin{bmatrix}
-1.29\\0\\1.37\\0.89\\2\\2\\10\\1\\0\\10
\end{bmatrix},
\end{align*}
and the corresponding charging control law is 
\begin{align*}
I_{k+1}=
\begin{bmatrix}
-2.263&2.263&0.938&0&0
\end{bmatrix}
\begin{bmatrix}
V_{b,k}\\V_{s,k}\\I_{k}\\\breve{r}\\u_{k-1}
\end{bmatrix},
\end{align*}
which is affine and easy to code and compute. While it is impossible to visualize the critical region partitioning in the five-dimensional space, one can make a cross-sectional view by fixing part of the parameters. With this idea, Figure~\ref{Fig:2D-plot} displays the partitioning of the critical regions on the two-dimensional $V_b$-$V_s$ plane when $I=2$~A, $\breve r=0.9$, and $\breve u=0$~A. It is seen that there are ten critical regions from this point of observation, with each one being a convex polygon.



Running the simulation, the eMPC charging control algorithm yields an optimal current profile as shown in Figure~\ref{Fig:current-simulation}, which distinguishes itself significantly from existing results. It is observed that this profile roughly includes three stages. Stage 1 features constant-current charging, which lasts for a relatively short period. Following it, Stage 2 sees the current decreasing at an approximately linear rate. The SoC increases fast during the two stages, and so does the terminal voltage $V$, as shown in Figures~\ref{Fig:SoC-simulation}-\ref{Fig:voltage-simulation}, respectively. When the charging continues in Stage 3, the magnitude of the current decreases at a faster rate overall, and the increase of SoC becomes slower. Meanwhile, it is seen that the rate of decrease is not uniform and alternates between fast and slow rythms. This is because the control law is seeking to achieve charging efficiency and constraint satisfaction simultaneously. If compared with the CC/CV charging, the optimal charging profile demonstrates more active regulation of the charging process, which believably can mitigate the effects of charging on the LiB cell's health more.
Figure~\ref{Fig:voltage-simulation} illustrates the actual terminal voltage based on the original nonlinear model and the one predicted by the linear models. One can observe a discrepancy between them, which results from the model approximation. However, the actual voltage always lies below the pre-set upper limit due to the reasons elaborated in Remark~\ref{Constraint-Satisfaction}. Figures~\ref{Fig:Vs-Vb-simulation} and~\ref{Fig:Vs-and-Vb} show $V_s-V_b$ and the profiles of $V_b$ and $V_s$, respectively. From all the figures, it is seen that every constraint is well satisfied throughout the charging process. These results indicate the efficacy of the proposed eMPC charging control algorithm to practical execution and its promise for health-aware charging.



It is understood that the eMPC algorithm approximates an NMPC formulated in~\eqref{Nonlinear-MPC-problem}, which involves no linearization and conducts online optimization. Figure~\ref{Fig:EMPC-VS-NMPC} compares the profiles of the charging current, SoC and terminal voltage generated by the two methods. It is seen that both lead to very close results. This indicates that the eMPC can almost reproduce the NMPC while offering much higher computational efficiency.

Recalling Remark~\ref{Remark:Complexity}, the online implementation of an eMPC mainly concerns search and evaluation of the PWA functions in the lookup table. One can use the sequential search method to retrieve the correct PWA function from the look-up table~\cite{tondel2003evaluation}. For the above simulation, if considering that the   search at every time step checks all critical regions in the worst case,  the computation would involve  around $840$ multiply-accumulate operations in total. Section~\ref{Sec:Evaluation-Efficiency} discusses further about the   running time. As for memory cost, the nine eMPCs here require a storage space for  roughly $9,756$ real numbers that encode all PWA functions and critical regions in the worst case. Our extensive simulations also illustrate   an arithmetic precision of three digits after the decimal point can assure sufficient control accuracy and performance.


\subsection{EKF-Based Output-Feedback Charging Control }\label{Sec:EKF-eMPC}
This paper focuses on the eMPC charging control design on the state-feedback assumption. The result can be easily modified to enable the more practical output-feedback charging control by adding an observer to carry out state estimation (see Figure~\ref{Fig:Algorithm-plot} and Remark~\ref{State-Feedback}). Here, let us investigate this extension by choosing the well-known EKF~\cite{fang2018nonlinear} as the observer and feeding the estimated states to the eMPC charging control algorithm. For the application of EKF, the model in~\eqref{new-state-space-equation} is considered and expanded to include process and measurement noise, which are both zero-mean white Gaussian and have covariances of $10^{-6} I$ and $9\times10^{-6}$, respectively. The other setting for simulation is the same as in Section~\ref{Sec:Case-Study}. Figures~\ref{Fig:EKF-current}-\ref{Fig:EKF-voltage} show the resultant profiles of the charging current, SoC and terminal voltage. Comparing them one-on-one with Figures~\ref{Fig:current-simulation}-\ref{Fig:voltage-simulation}, one can find out some difference between them due to the noise and state estimation errors, which nonetheless is very small. Figure~\ref{Fig:EKF-Vs-Vb} displays $V_s-V_b$ slightly fluctuating around the pre-set constraint, still because of the estimation errors. While the constraint is not fully satisfied here, the violation is at a quite minor level without making a concern. Figure~\ref{Fig:EKF-Vs-and-Vb} depicts the profiles of the actual $V_b$ and $V_s$, which are only marginally different from their counterparts in Figure~\ref{Fig:Vs-and-Vb}. The results highlight that the proposed eMPC charging control algorithm can be implemented efficiently and effectively in the output-feedback manner in practice.

\begin{figure*} [t]
\centering 
\subfigure[Charging current profile]
{\includegraphics[trim = {0 0 0 0}, clip, width=0.3\textwidth]{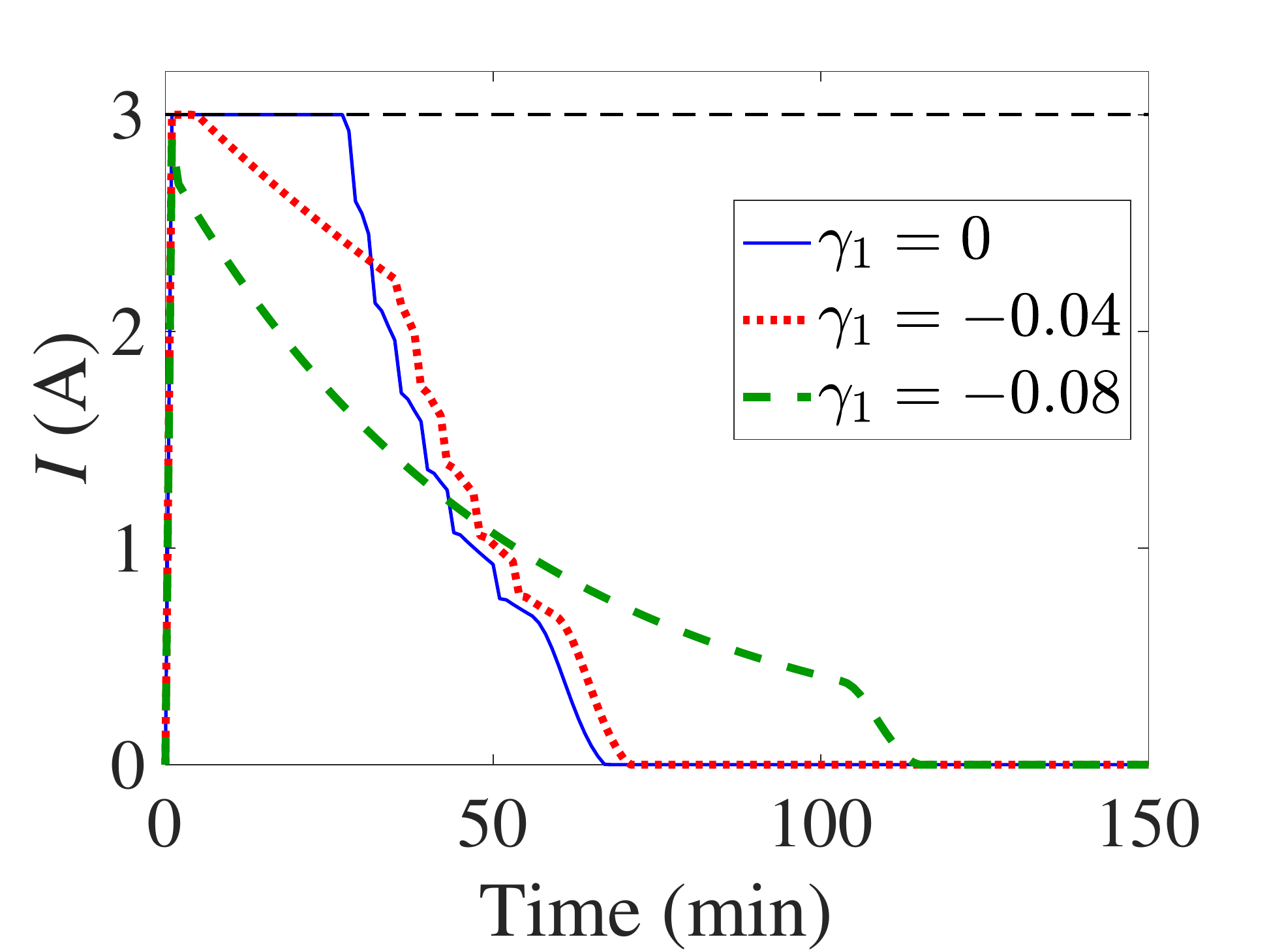}\label{Fig:VsVb-current}} 
\hspace{0in} 
\subfigure[SoC profile]
{\includegraphics[trim = {0 0 0 0}, clip, width=0.3\textwidth]{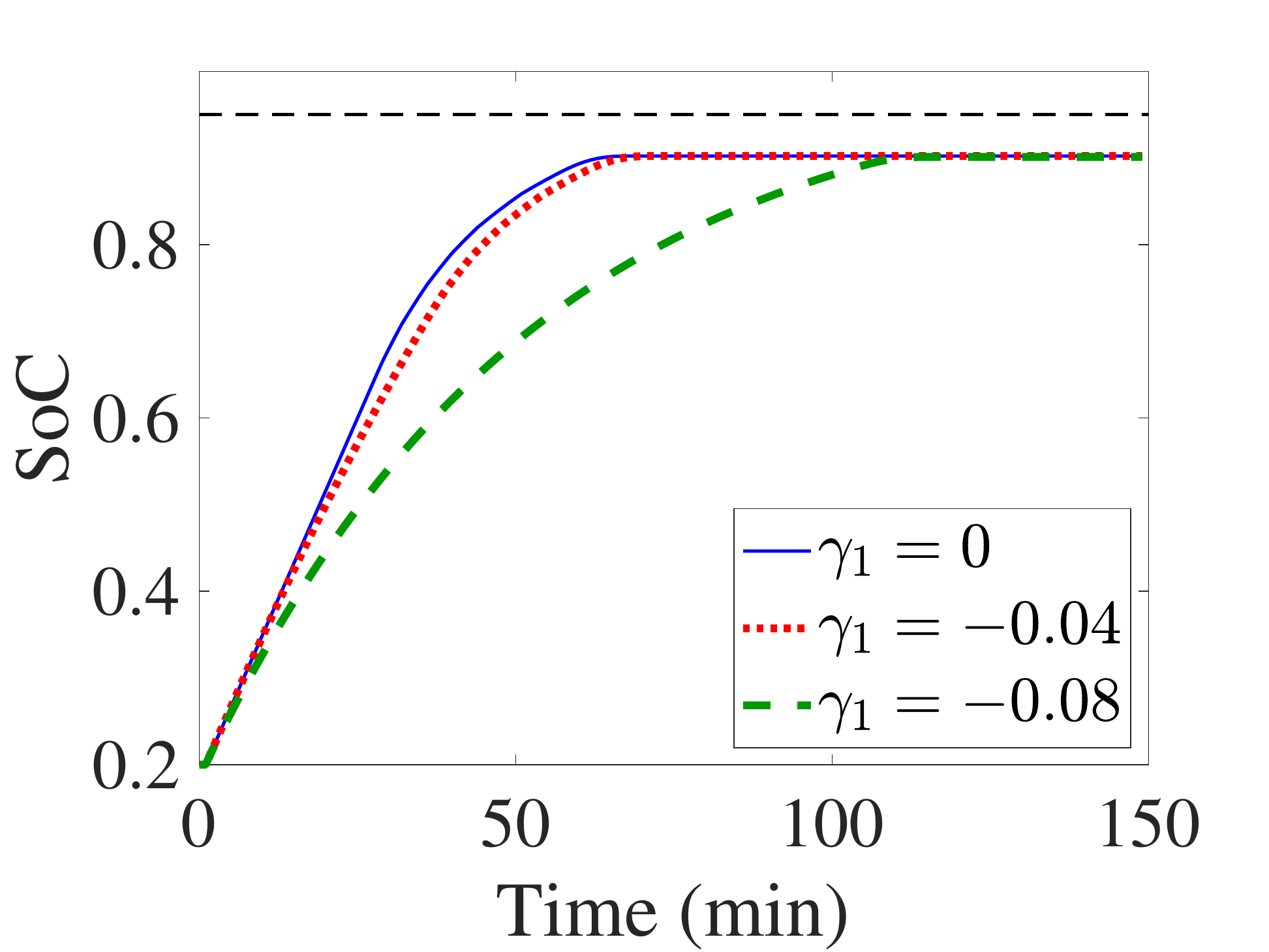}\label{Fig:VsVb-SoC}}
\subfigure[Voltage profile]
{ \includegraphics[trim = {0 0 0 0}, clip, width=0.3\textwidth]{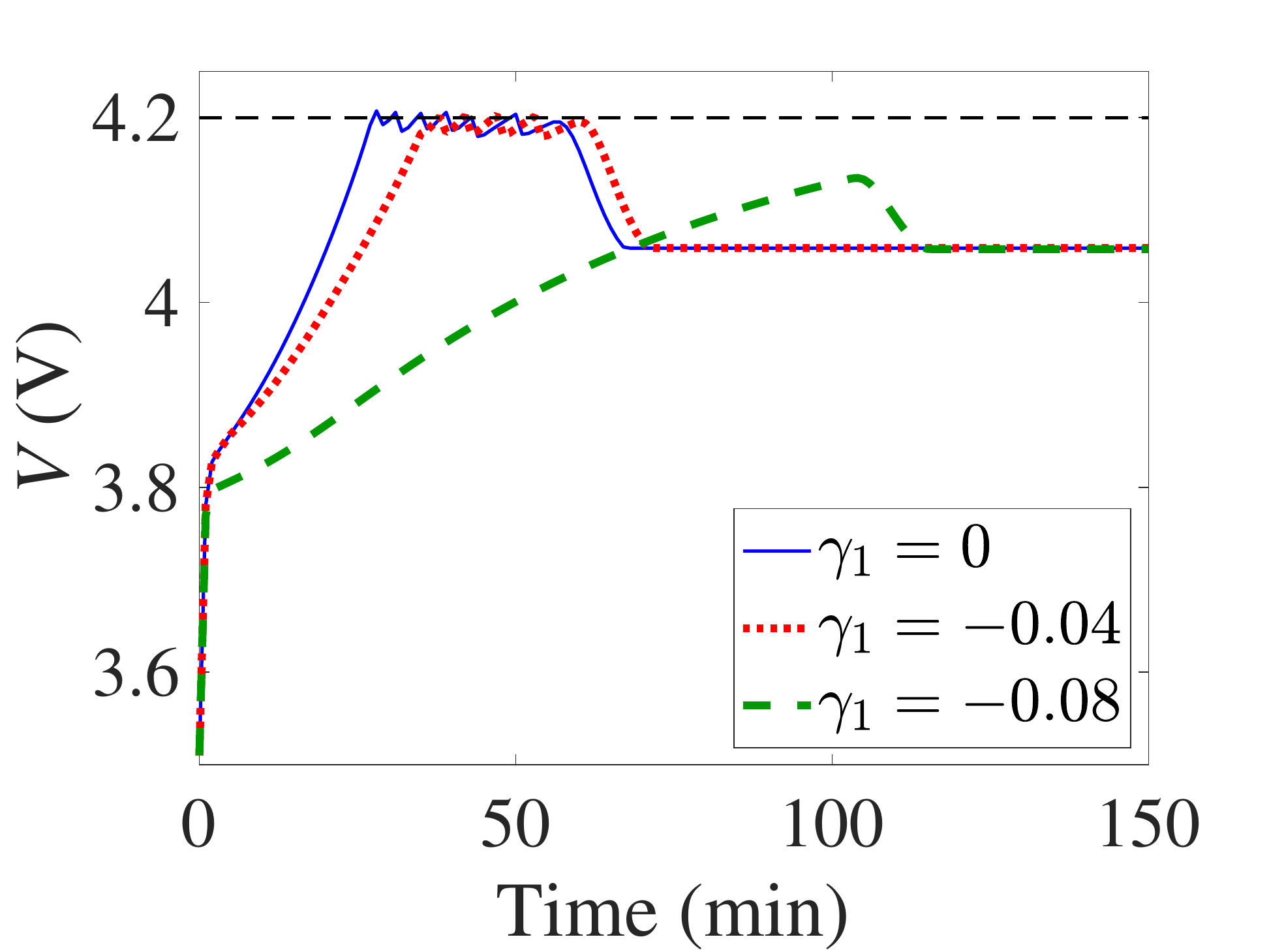}\label{Fig:VsVb-voltage}} 
\caption{Charging control under different constraints on $V_s-V_b$.} 
\label{Fig:VsVb-constraints}
\vspace{-5mm}
\end{figure*}

\begin{figure} [t]
\centering 
\includegraphics[trim = {0 0 0 0}, clip, width=0.3\textwidth]{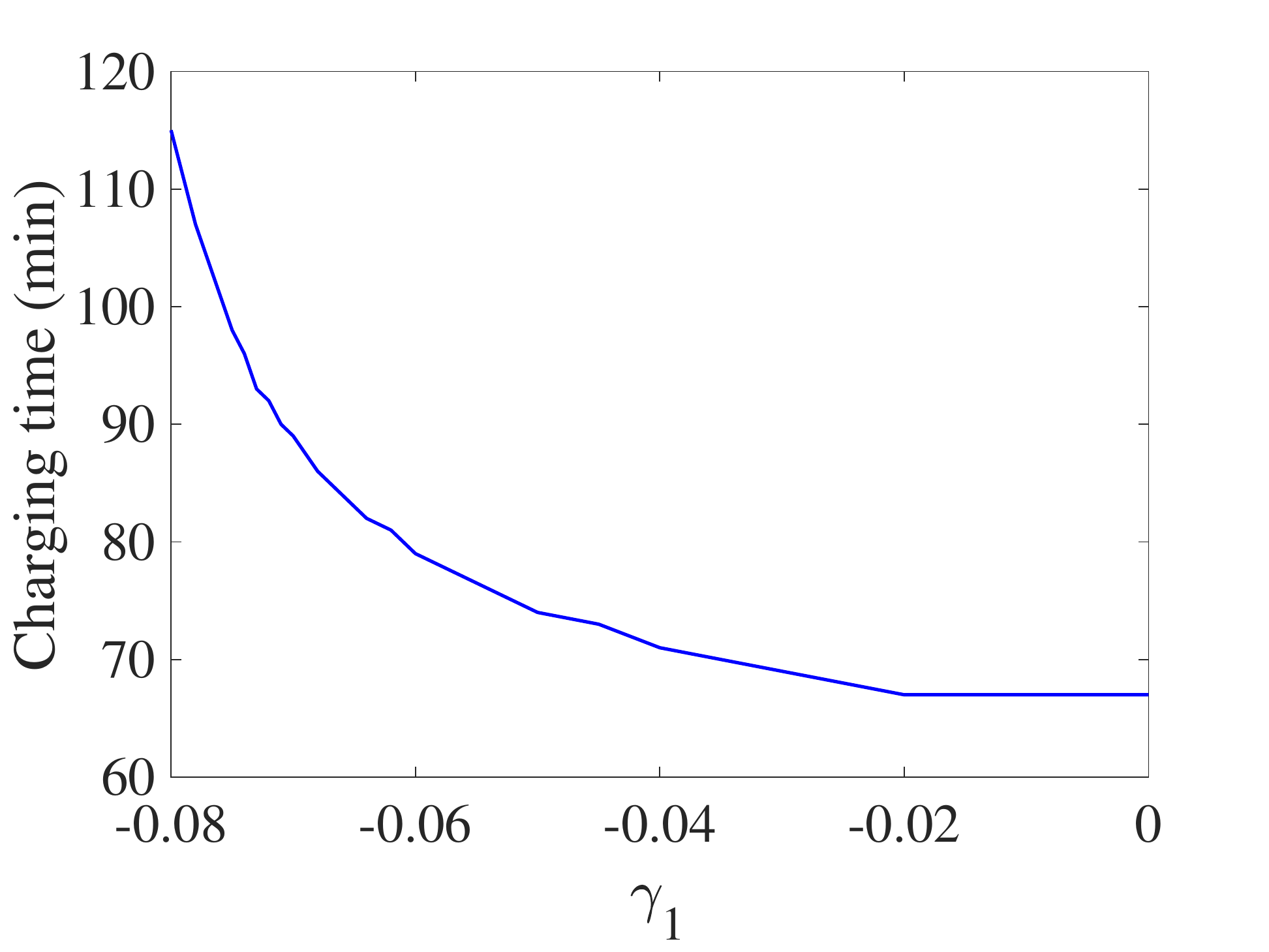}
\caption{Charging time versus $\gamma_1$ under $\gamma_2=0.08$.} 
\label{Fig:Charging-Time-Gamma}
\vspace{-5mm}
\end{figure}

\subsection{Effects of Changing Constraints and Horizon Parameters}\label{Sec:Design-Parameters}

Constraints and horizon parameters play a vital role in MPC-based optimal charging design. This section examines how the constraint~\eqref{Vs-Vb-gamma12} and the prediction, control and constraint horizons affect the proposed algorithm. 

Recall that the constraint~\eqref{Vs-Vb-gamma12} is determined by $\gamma_1$ and $\gamma_2$. Different choices of them will lead to different levels of health consciousness. Here, fix $\gamma_2$ at $0.08$. Then let $\gamma_1$ take $0$, $-0.04$ and $-0.08$, respectively, yielding increasingly stricter restrictions on $V_s-V_b$. The other parameters follow the ones in Section~\ref{Sec:Case-Study}. Figure~\ref{Fig:VsVb-constraints} summarizes the simulation results in this setting. Figure~\ref{Fig:VsVb-current} illustrates that, when $\gamma_1$ decreases, the ``head'' of the charging current profile lowers, and the ``tail'' raises and lengthens, leading to a longer charging time. This results from the constraint $V_s-V_b$ becoming more restrictive to enforce stronger health protection in charging. The corresponding SoC and terminal voltage profiles are plotted in Figures~\ref{Fig:VsVb-SoC}-\ref{Fig:VsVb-voltage}. Both of them rise more slowly when $\gamma_1$ switches from $0$ to $-0.04$ and then $-0.08$. From these plots, one can observe that the stronger health protection in charging can be compromised by the longer charging time. This is verified by Figure~\ref{Fig:Charging-Time-Gamma}, which gives an illustration of the charging time versus $\gamma_1$ when $\gamma_2$ is fixed at $0.08$. A major implication is that a practitioner will need to select $\gamma_1$ and $\gamma_2$ to strike a balance between charging time and battery health, depending on the considered application. A simulation-based trial-and-error procedure can be used for the search.

\begin{figure*} [t]
\centering 
\subfigure[Charging current profile]
{\includegraphics[trim = {0 0 0 0}, clip, width=0.3\textwidth]{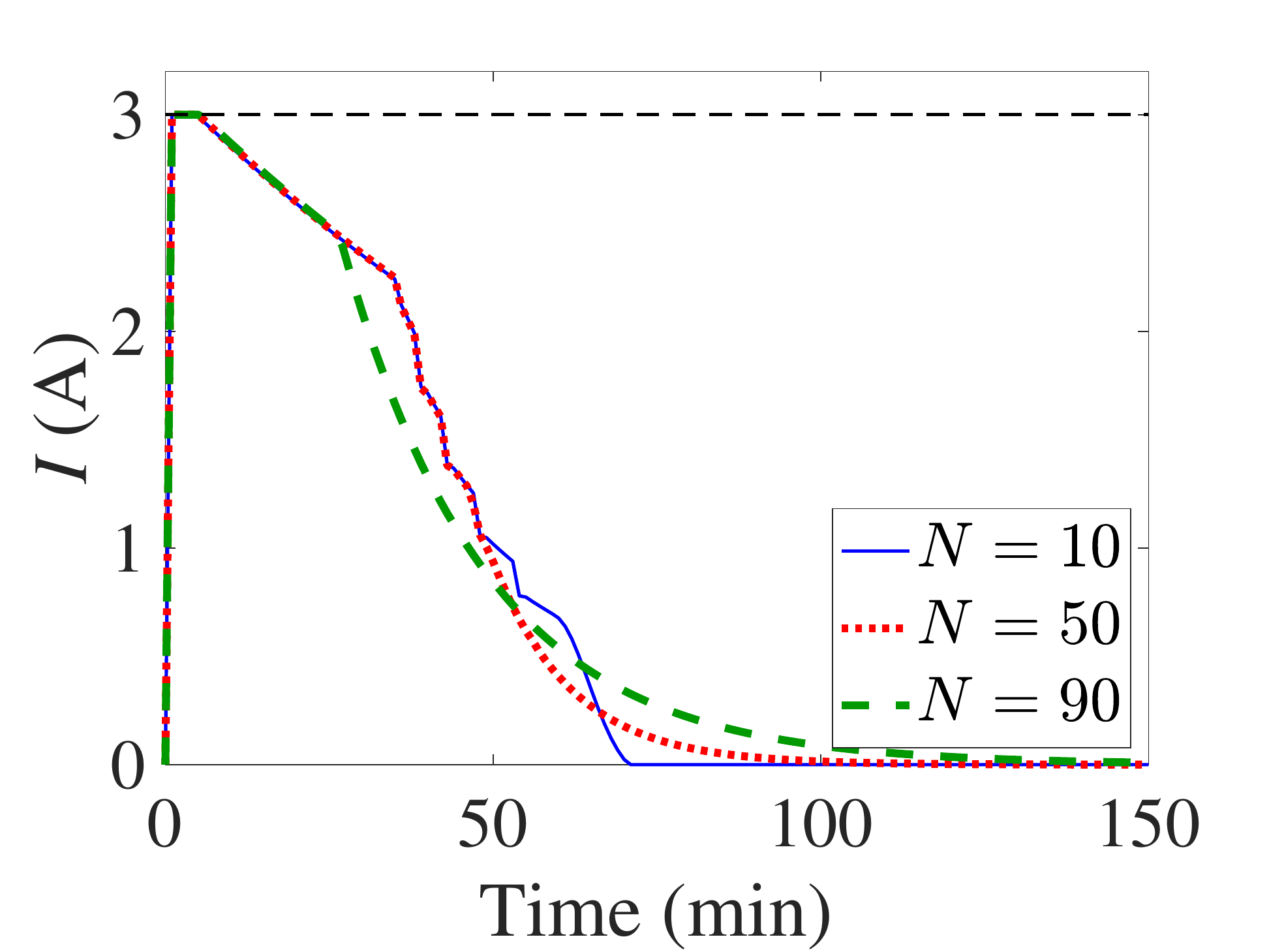}\label{Fig:current-prediction}} 
\hspace{0in} 
\subfigure[SoC profile]
{\includegraphics[trim = {0 0 0 0}, clip, width=0.3\textwidth]{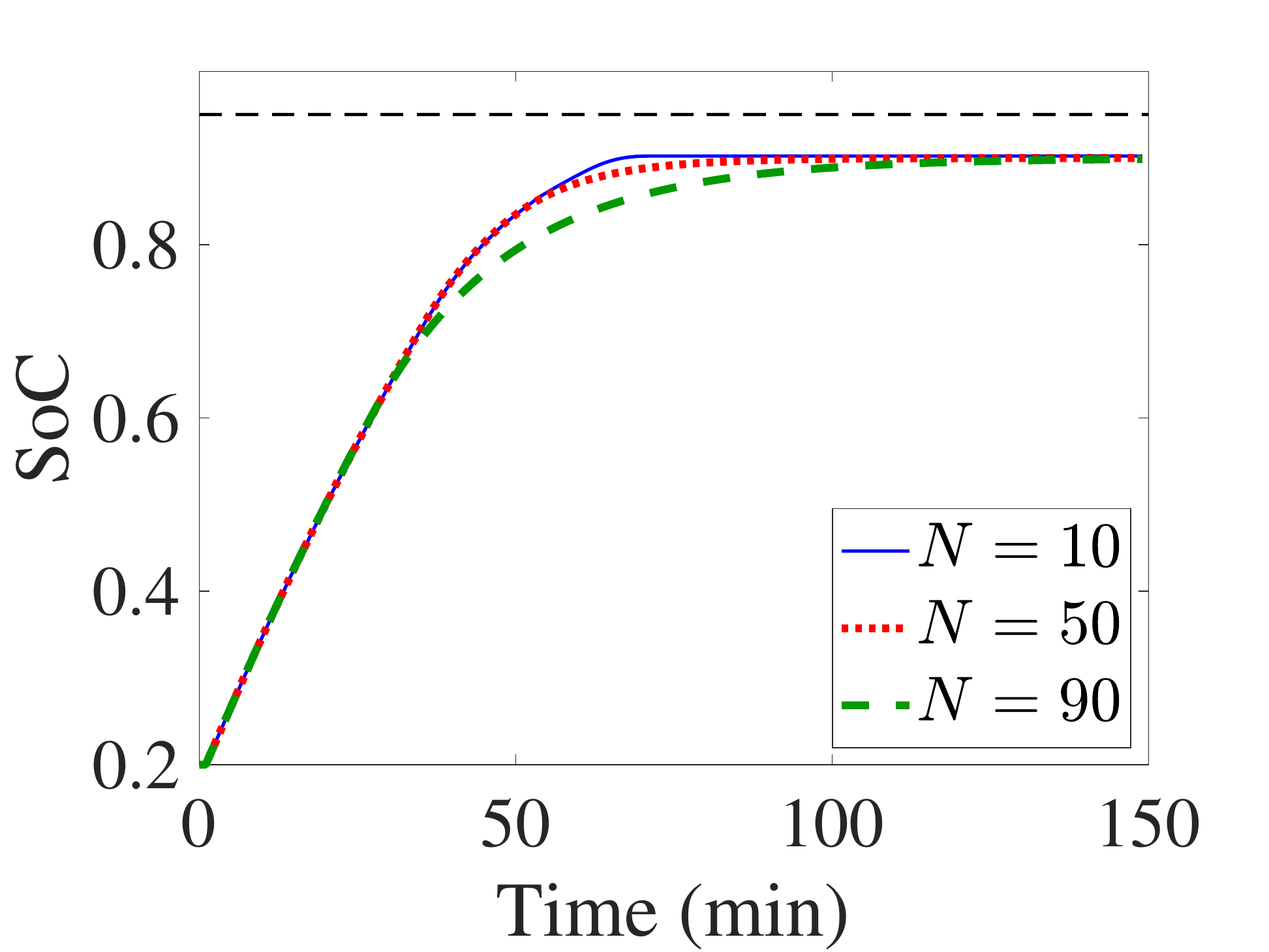}\label{Fig:SoC-prediction}}
\hspace{0in} 
\subfigure[Voltage profile]
{ \includegraphics[trim = {0 0 0 0}, clip, width=0.3\textwidth]{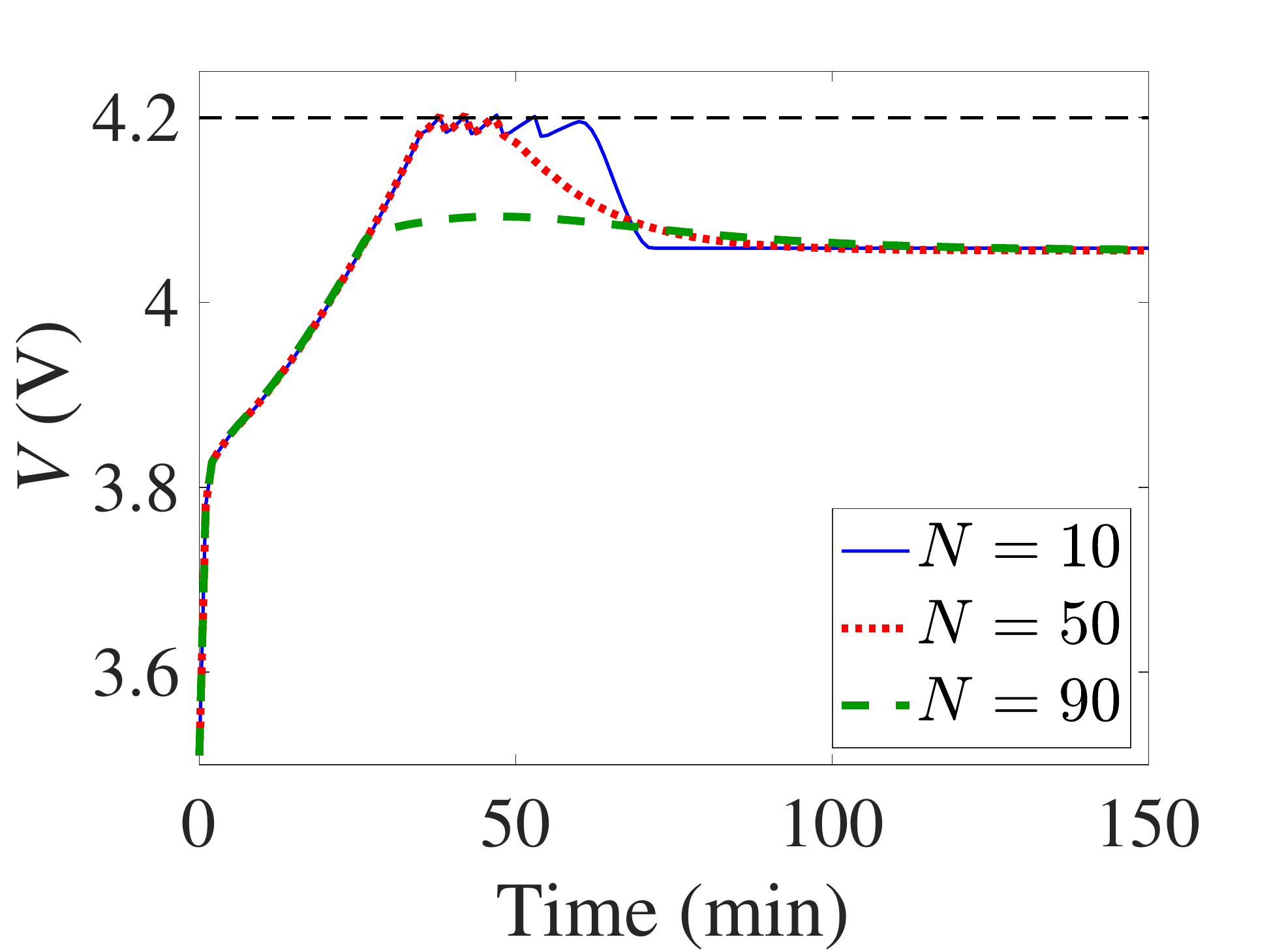}\label{Fig:voltage-prediction}} 
\caption{Charging control when the prediction horizon $N=10,50,90$.} 
\label{Fig:variable-prediction}
\vspace{-5mm}
\end{figure*}

Now, let us vary the horizon parameters and assess their influence on the charging results. To this end, it is beneficial to focus on only one horizon parameter at one time and have all the other parameters remain the same as in Section~\ref{Sec:Case-Study}. First, consider the prediction horizon $N$, and let it take $10$, $50$ and $90$, respectively. Figure~\ref{Fig:variable-prediction} depicts the charging current, SoC and terminal voltage profiles obtained for different choices of $N$. Here, it is seen that an increasing $N$ would make the charging process smoother and slower. This is because the optimization now is about a cost function defined and evaluated over a longer-term future, the solution of which will hence lead to less aggressive short-term control. It is also noteworthy that the charging current profiles differ only moderately despite different $N$. The reason lies in the slow dynamics of LiBs, which traces its origin to the fact that the two eigenvalues of $A$ are either at or close to the origin. Because of this, if comparing the eMPC with the original NMPC running with a very large prediction horizon, one can also find out that they lead to close performance. This implies that a relatively small $N$ can be a safe choice.

\begin{figure*} [t]
\centering 
\subfigure[Charging current profile]
{\includegraphics[trim = {0 0 0 0}, clip, width=0.3\textwidth]{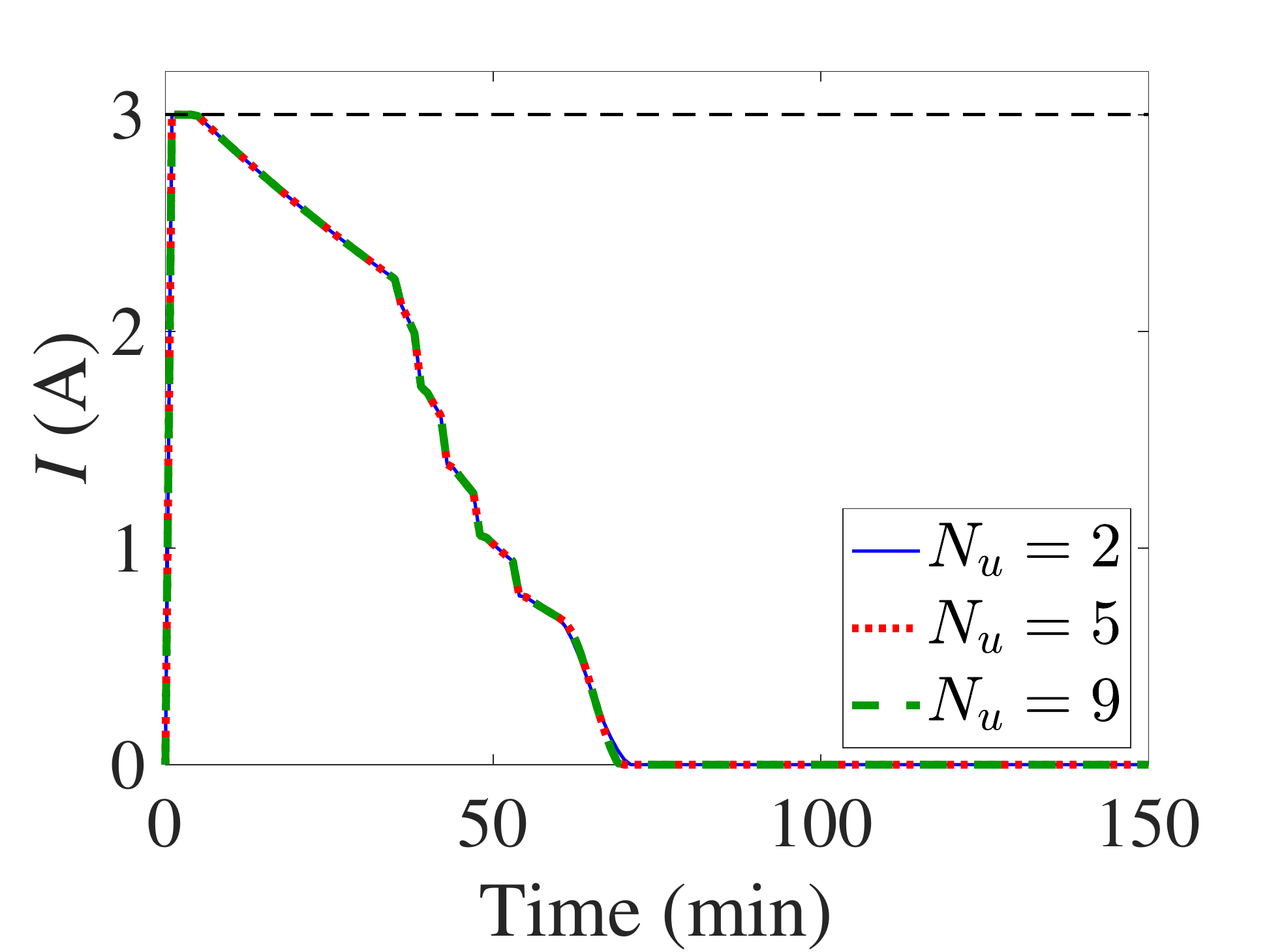}\label{Fig:current-control}} 
\hspace{0in} 
\subfigure[SoC profile]
{\includegraphics[trim = {0 0 0 0}, clip, width=0.3\textwidth]{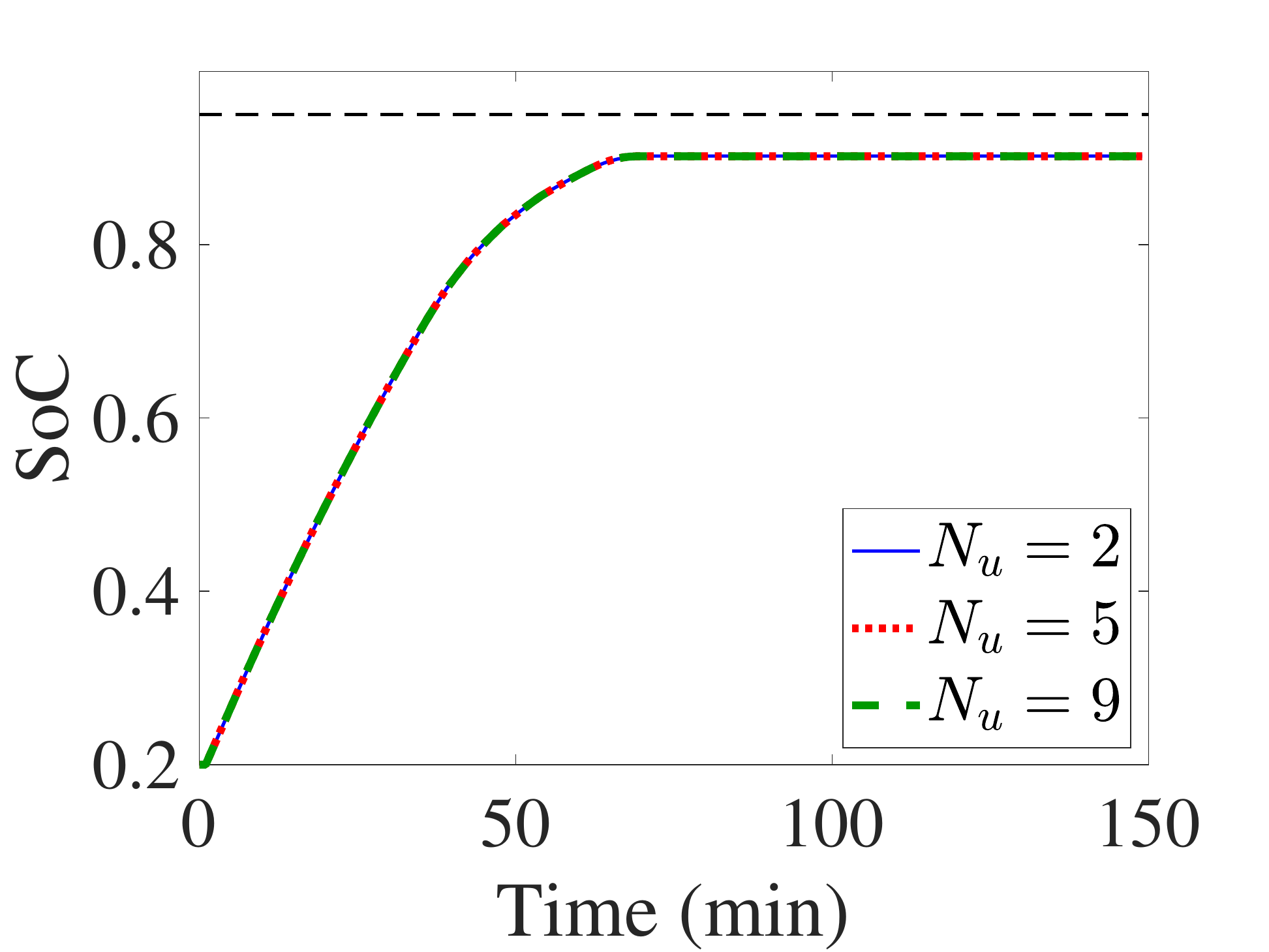}\label{Fig:SoC-control}}
\hspace{0in} 
\subfigure[Voltage profile]
{ \includegraphics[trim = {0 0 0 0}, clip, width=0.3\textwidth]{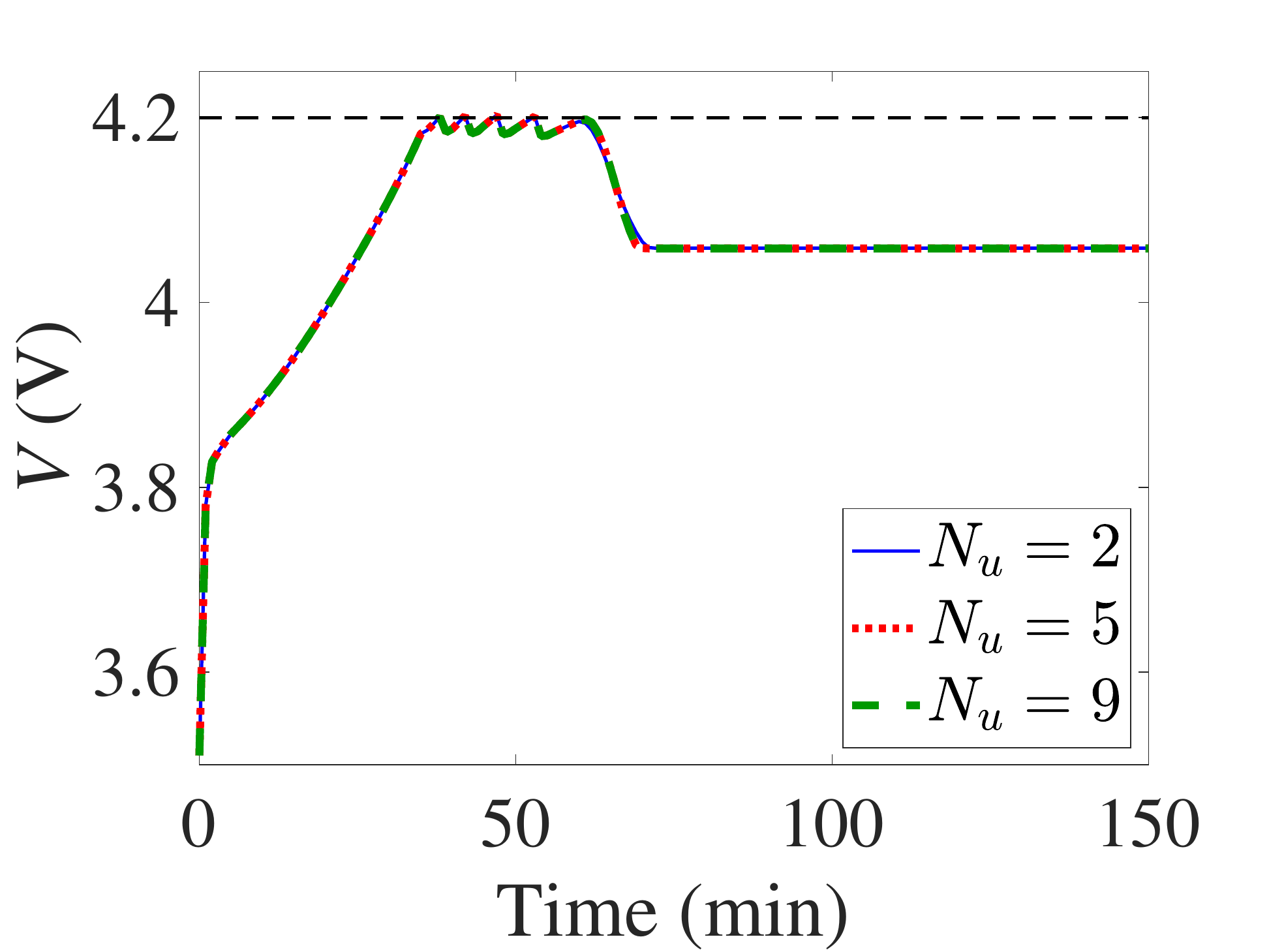}\label{Fig:voltage-control}} 
\caption{Charging control when the control horizon $N_u=2,5,9$.} 
\label{Fig:variable-control}
\vspace{-5mm}
\end{figure*}

\begin{figure*} [t]
\centering 
\subfigure[Charging current profile]
{\includegraphics[trim = {0 0 0 0}, clip, width=0.3\textwidth]{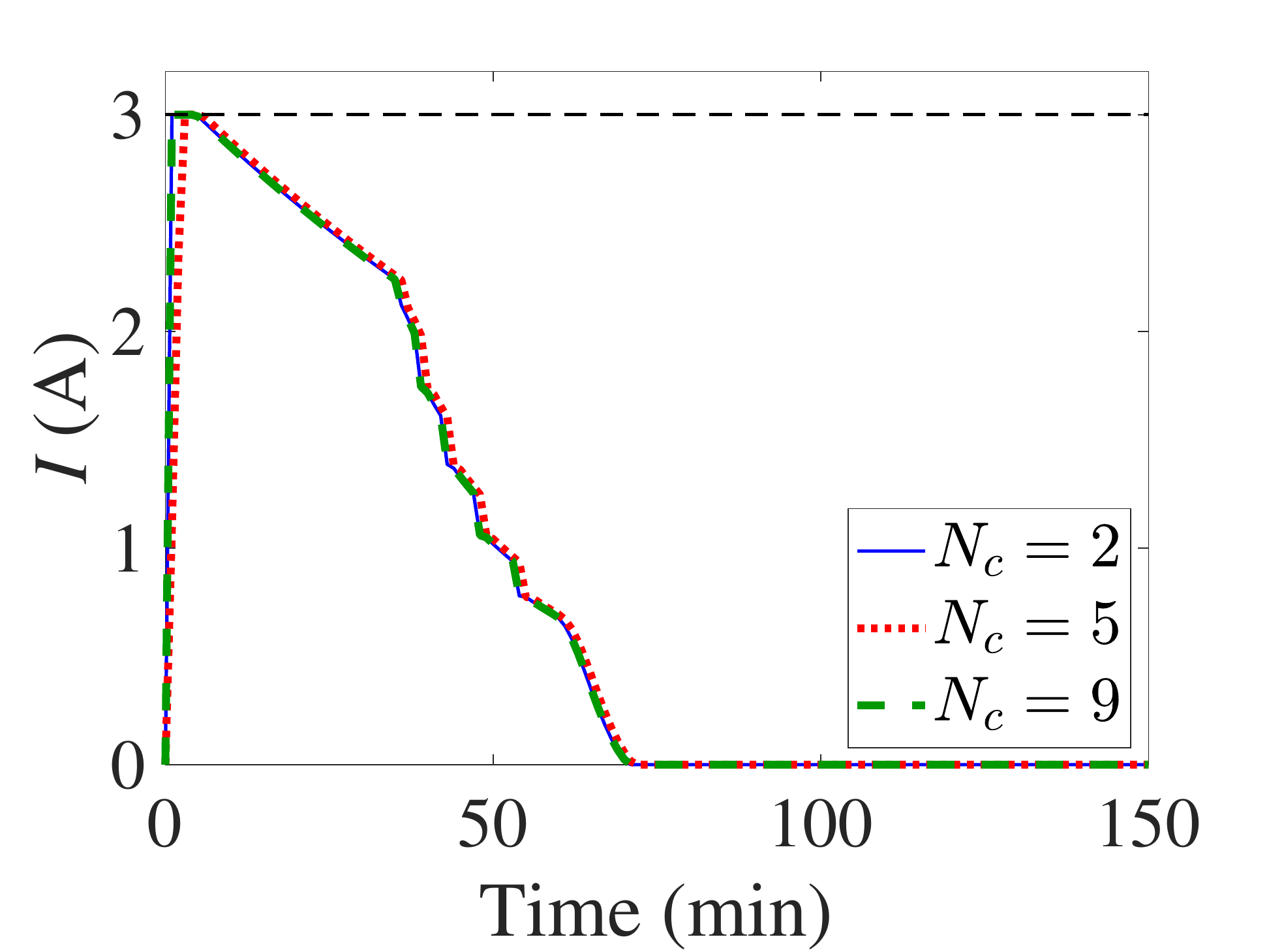}\label{Fig:current-constraint}} 
\hspace{0in} 
\subfigure[SoC profile]
{\includegraphics[trim = {0 0 0 0}, clip, width=0.3\textwidth]{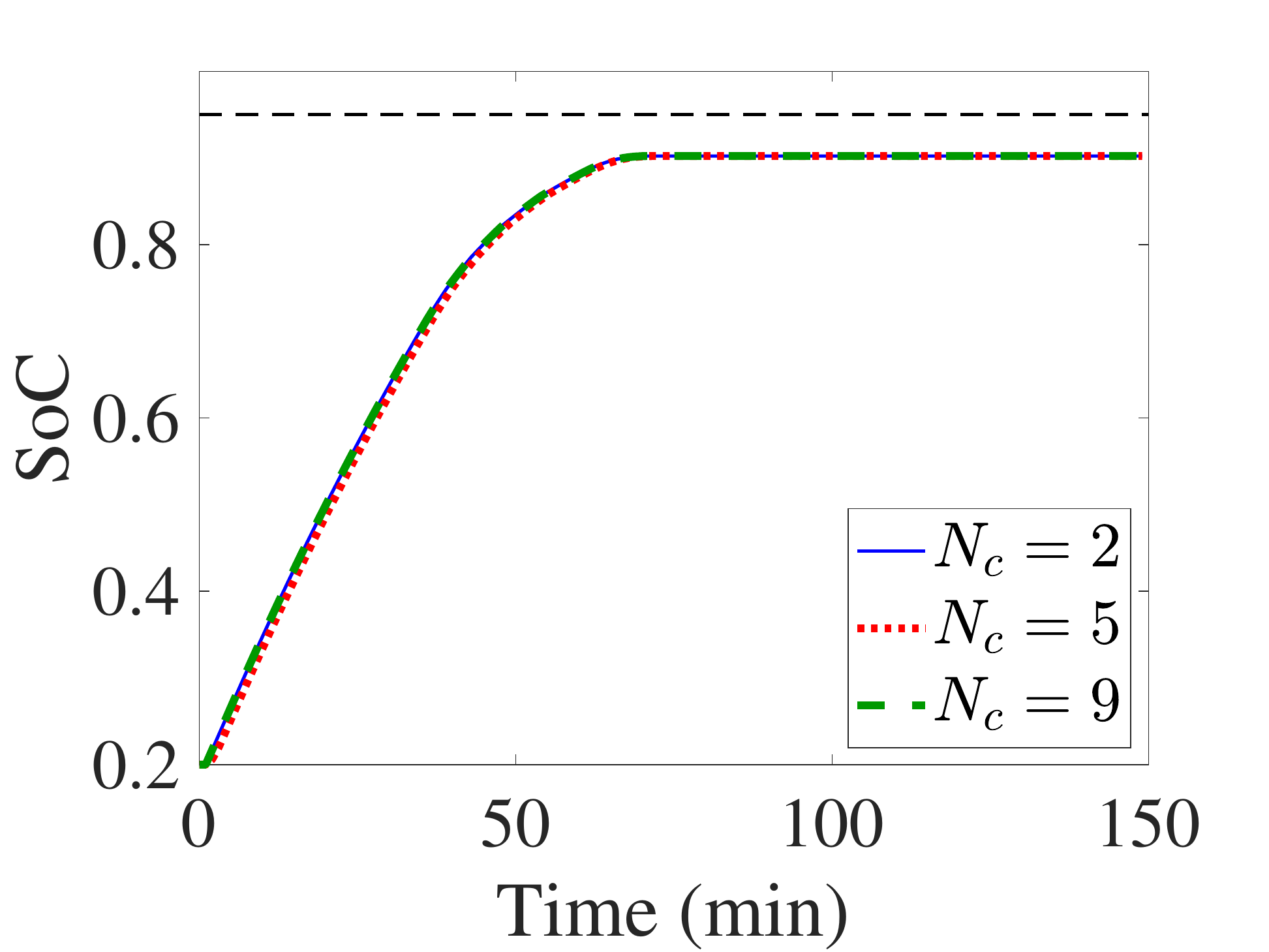}\label{Fig:SoC-constraint}}
\hspace{0in} 
\subfigure[$V_s-V_b$]
{ \includegraphics[trim = {0 0 0 0}, clip, width=0.3\textwidth]{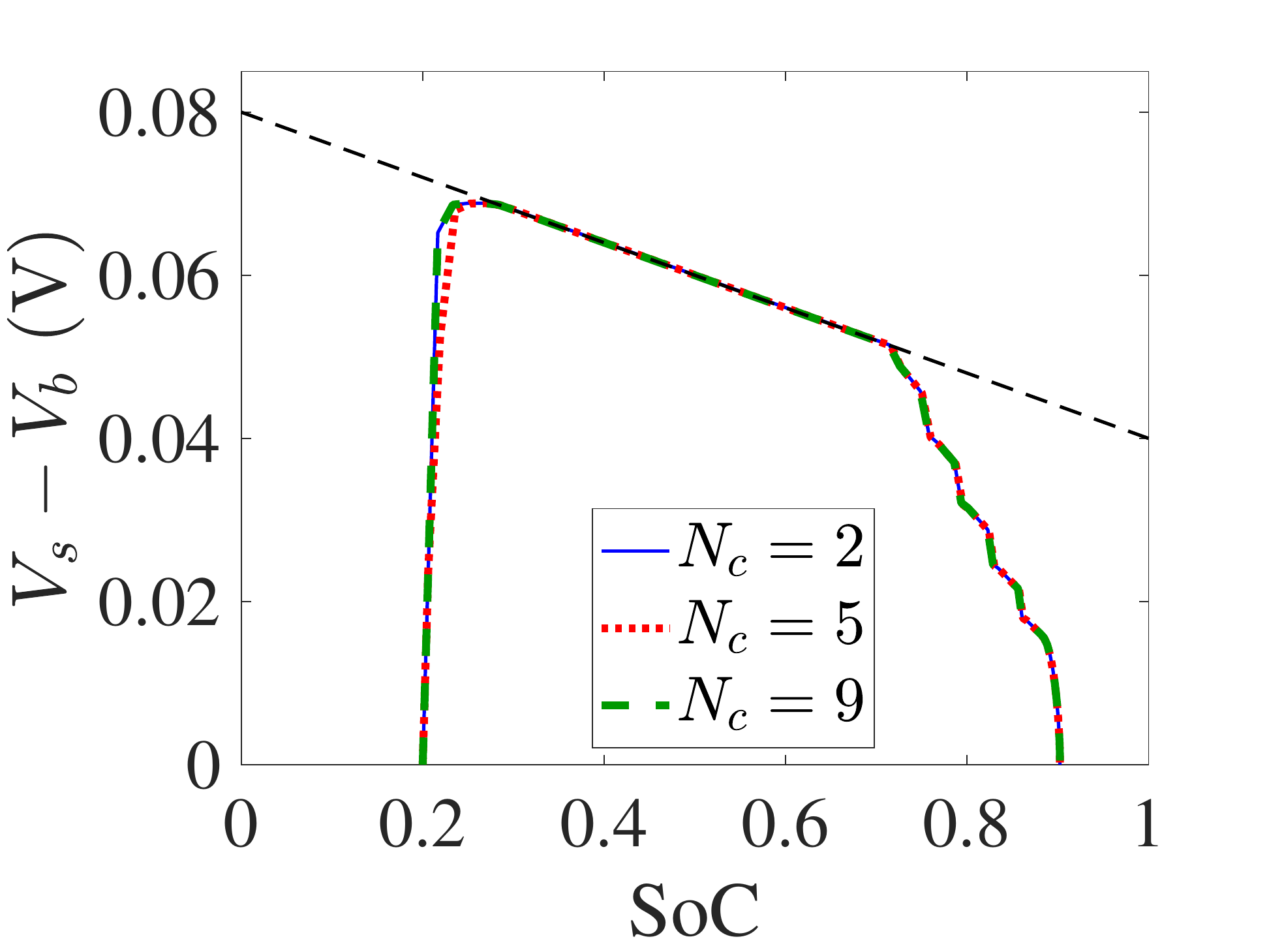}\label{Fig:Vs-Vb-constraint}} 
\caption{Charging control when the constraint horizon $N_c=2,5,9$ for the constraint on $V_s-V_b$.} 
\label{Fig:variable-constraint}
\vspace{-5mm}
\end{figure*}

Next, look at the control horizon $N_u$, which represents the number of charging moves to be optimized. Usually, $N_u \ll N$, and here, use $N_u =2,~5$ and $9$, respectively. Figure~\ref{Fig:variable-control} shows that almost the same results are acquired for different choices of $N_u$, even though the charging becomes a little faster when $N_u$ increases. This is also largely due to the slow dynamics of LiBs. As a result, changing $N_u$ by a scale of several minutes will produce little change to $\mathrm \Delta u_k$. An additional reason is that the enforced constraints further suppress the variation of $\mathrm \Delta u_k$. This observation suggests that a small $N_u$ would be sufficient in practice, which will also reduce the computation advantageously as it leads to fewer critical regions. As an extra benefit, a smaller $N_u$ implies fewer control variables in the mpQP computation and brings a parameter space divided by fewer critical regions, thus yielding greater computational efficiency for both offline and online computation. Finally, let the constraint horizon $N_c$ for the constraint in~\eqref{Vs-Vb-gamma12} vary among $2$, $5$ and $9$, respectively. Figure~\ref{Fig:variable-constraint} shows almost identical results for the different choices of $N_c$. To see why, one can examine the dynamics of $V_s-V_b$, which, by~\eqref{state-space-equation}, is given by
\begin{align*}
\dot{V}_s-\dot{V}_b
= -
\frac{C_b+C_s}{C_bC_s(R_b+R_s)}
\left(V_s-V_b \right) + 
\frac{R_bC_b-R_sC_s}{C_bC_s(R_b+R_s)}I .
\end{align*}
Since the first term of the right-hand side is larger than the second one by at least two orders of magnitude, the current $I$ has only marginal influence on the dynamics of $V_s-V_b$. The current profile hence will not change much even if $N_c$ changes. As is with the case for $N_u$, this phenomenon allows us to use a small $N_c$ for the constraint on $V_s-V_b$, which can assure control performance while promoting faster computation. 

Although the horizon parameter selection for the proposed eMPC charging control algorithm would require some empirical optimization in practice, the following suggestions are offered summarizing the above: 
\begin{enumerate}

\item One can select an $N$ such that the prediction horizon is about ten minutes. 

\item It is advisable to let $N_u,N_c \ll N$. Small $N_u$ and $N_c$ will lead to charging current profiles similar to those resulting from large $N_u$ and $N_c$ but can reduce online computational costs. The above results show that it is sufficient to set $N_u,N_c=2$ if $N=10$.

\item Trial-and-error tuning based on simulations can help a practitioner determine the best horizon parameters for specific applications. 
\end{enumerate}

\begin{table}[t]\centering 
\begin{threeparttable}
\caption{Comparison of computational time.}
\begin{tabular}{cccc}
\toprule%
Method& $(N,N_u,N_c)$ &Time (s)& Case \\
\midrule
eMPC & (10,2,2)& 0.42 & Figures 5-6, 10-12 \\
& (50,2,2)& 0.42 & Figure 10\\
& (90,2,2)& 0.43 & Figure 10\\
& (10,5,2)& 0.42 & Figure 11\\
& (10,9,2)& 0.42 & Figure 11\\
& (10,2,5)& 0.42 & Figure 12\\
& (10,2,9)& 0.42 & Figure 12\\
\midrule
NMPC & (10,2,2)& 6.45& Figure 6\\
\bottomrule
\end{tabular}
\label{Tab:Computation-Time}
\end{threeparttable}
\vspace{-2mm}
\end{table}

\subsection{Evaluation of Computational Efficiency}\label{Sec:Evaluation-Efficiency}

Finally, let us conclude the simulation by comparing the running time of the proposed eMPC and the NMPC charging algorithms based on~\eqref{Nonlinear-MPC-problem}. To this end, a series of simulations were run on a Macbook Pro equipped with a 2.3 GHz Inter Core i5, 8 Gb RAM and MATLAB R2018b. Here, the evaluation of eMPC used different horizon parameter settings as considered in Section~\ref{Sec:Design-Parameters}. For each simulation, the entire control run comprised 150 time steps representing a charging session of 2.5 hours. To make a fair assessment, 20 simulation runs were conducted in each case, and the average running time was calculated. Table~\ref{Tab:Computation-Time} summarizes the computational time for all cases. It shows that eMPC on average takes no more than $1\over 15$ time of NMPC and  its computation for each time step is only about 3 milliseconds. The results highlight the computational superiority of eMPC over NMPC and show its promise for real-time charging management.

\begin{figure*} [t]
\centering 
\subfigure[Voltage profile under $\gamma_1=0$]
{\includegraphics[trim = {0 0 0 0}, clip, width=0.3\textwidth]{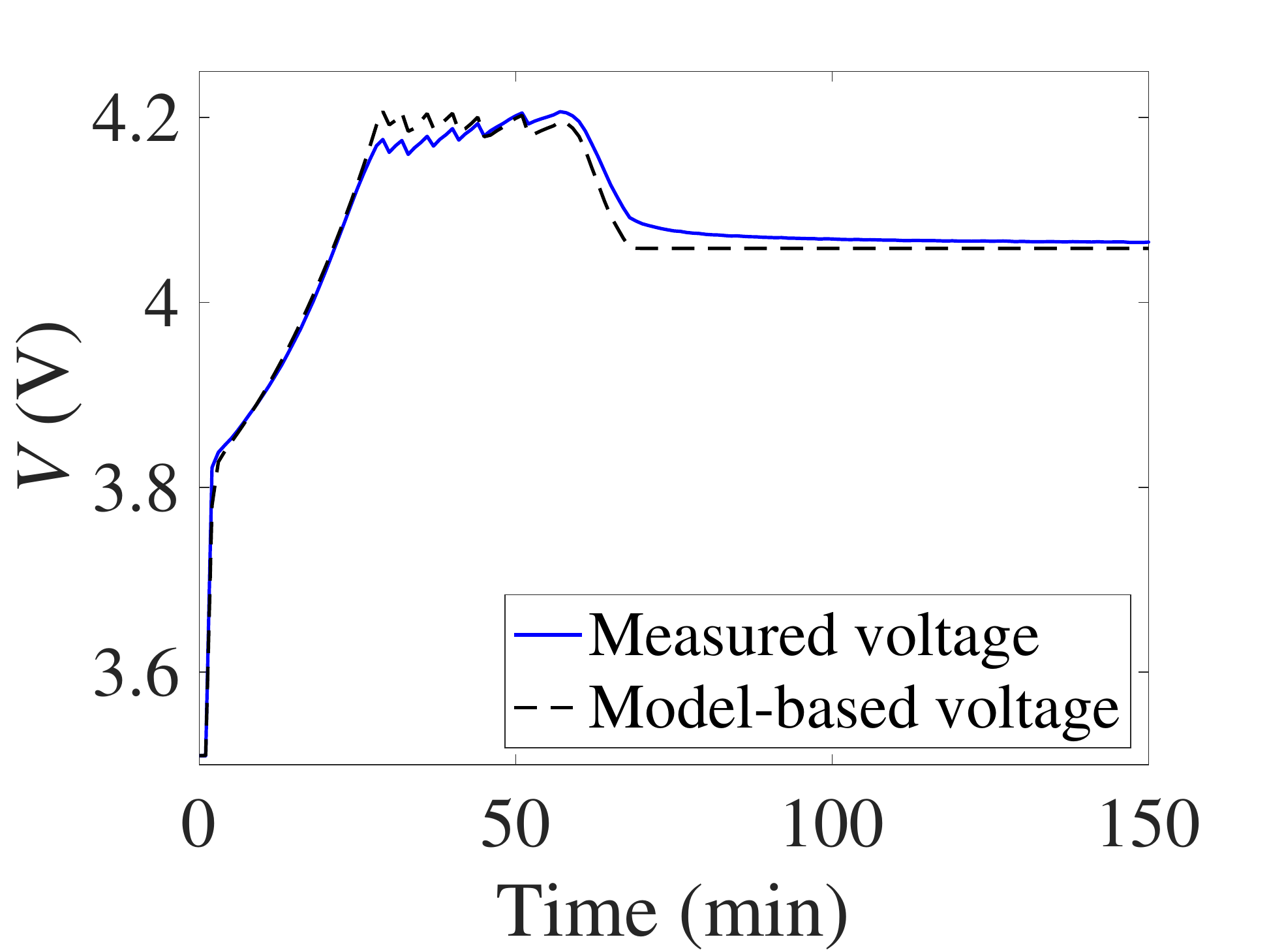}\label{Fig:flat-measurement}} 
\hspace{0in} 
\subfigure[Voltage profile under $\gamma_1=-0.04$]
{\includegraphics[trim = {0 0 0 0}, clip, width=0.3\textwidth]{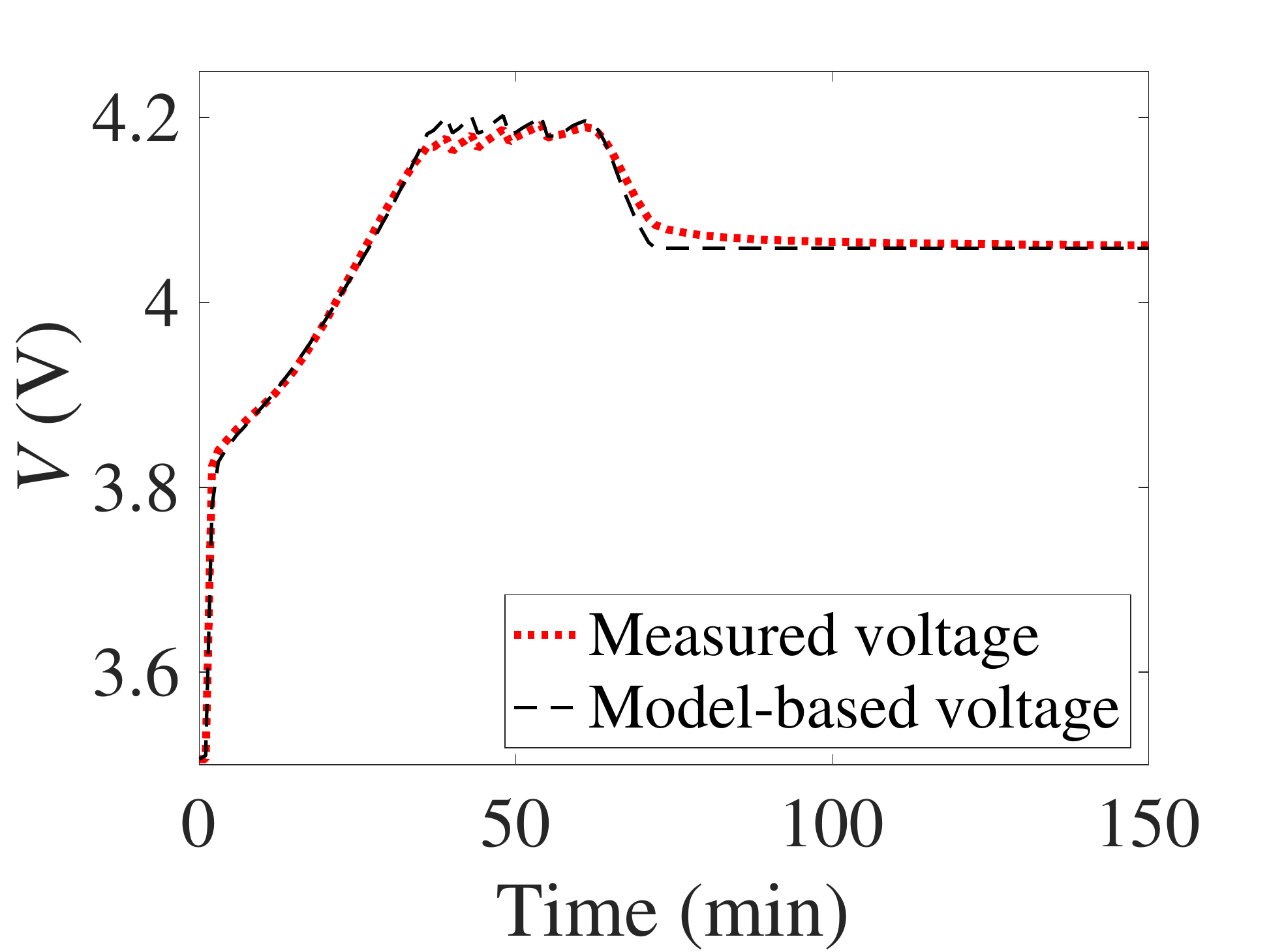}\label{Fig:medium-measurement}} 
\hspace{0in} 
\subfigure[Voltage profile under $\gamma_1=-0.08$]
{\includegraphics[trim = {0 0 0 0}, clip, width=0.3\textwidth]{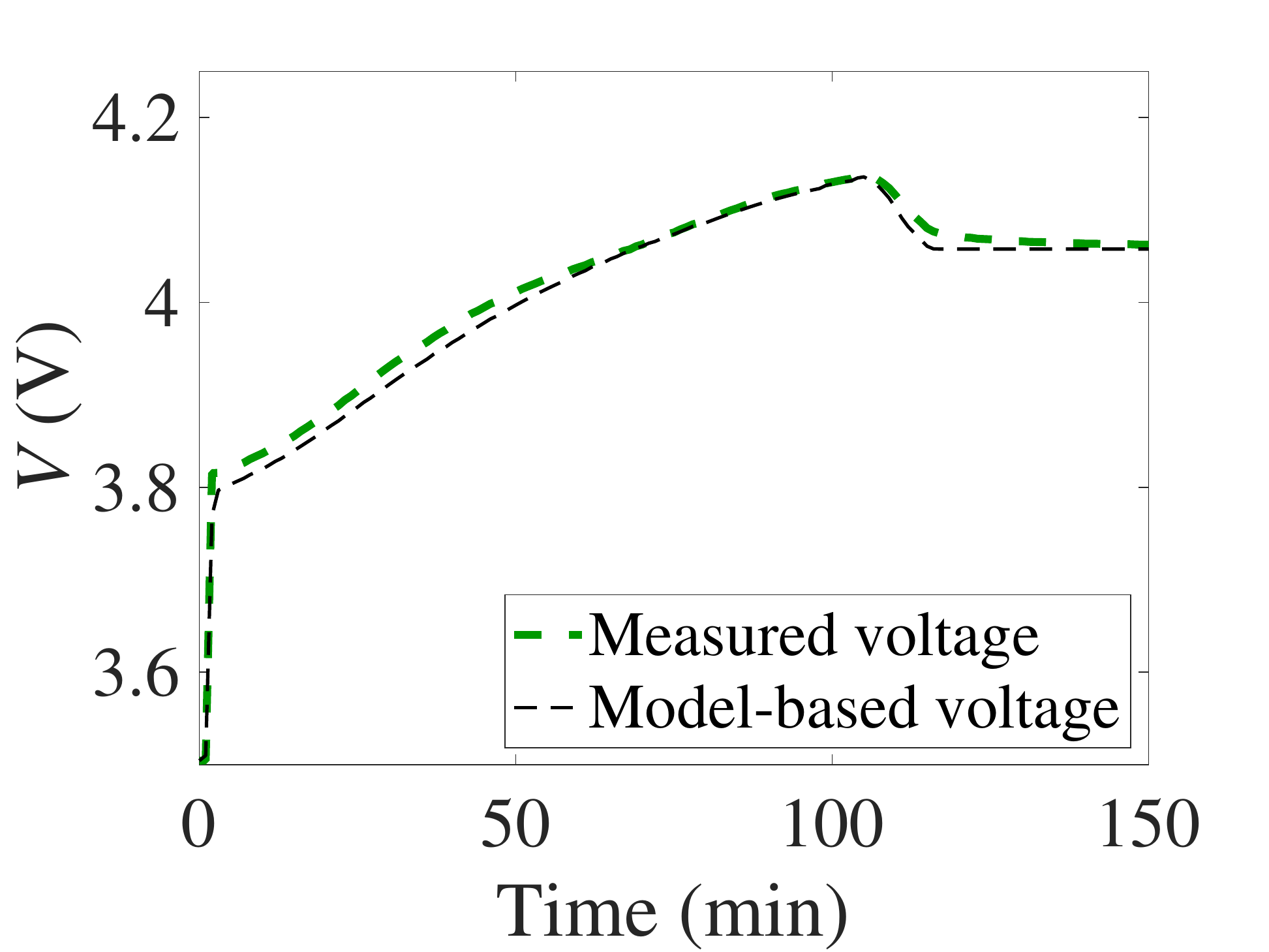}\label{Fig:steep-measurement}} 
\caption{Experimental results based on the proposed charging strategies under different $V_s-V_b$ constraints.} 
\label{Fig:Experimental-Results}
\vspace{-5mm}
\end{figure*}


\section{Experimental Validation}\label{Sec:Experimental-Validation}

The experimental validation was conducted using a PEC\textsuperscript{\textregistered} SBT4050 battery tester on a Panasonic NCR18650B LiB cell with a rated capacity of 3 Ah. The tester can support charging/discharging with arbitrary current-, voltage-, and power-based loads (up to 40 V and 50 A). It runs with a specialized server, which is used to prepare, configure and monitor a test through the associated software application LifeTest\textsuperscript{TM}. 
While such a tradeoff makes it difficult to fully reveal the strengths and utility of the proposed charging control algorithm, it is necessary for now to live with the facility's limitation and still allows useful results to be collected. 


The objective and setting of the experiment follow those for the simulation in Section~\ref{Sec:Case-Study}. The cell's parameters are also the ones used in the simulation and given in Table~\ref{Tab:Model-Parameters}. Hence, the experiment directly used the charging current profiles plotted in Figure~\ref{Fig:VsVb-current} to charge the cell, which represent optimal health-conscious charging designs when the constraint~\eqref{Vs-Vb-gamma12} takes different parameters. Figures~\ref{Fig:flat-measurement}-\ref{Fig:steep-measurement} illustrate the measured terminal voltage for each charging current profile and make a comparison with the model-based prediction, where a close match between them is observed. The experiment verifies that the proposed algorithm can be well used for practical charging control. It is noted that battery life cycle testing is desired to further evaluate its role in mitigating health degradation, which will be part of our future work.


Here, due to the limitation of the tester, a charging profile was computed offline and then was uploaded to the tester to charge the cell in an open-loop control manner. Even though this would lead to certain performance loss, the results shown above still illustrate the effectiveness of the charging control. It is believed that better control performance will be achieved if applying the algorithm to closed-loop control, which will be pursued in our future work.

\section{Conclusion}\label{Sec:Conclusion}

Charging control is essential for the health and safety of LiBs, which sees an ever-growing importance as they are becoming one of the most ubiquitous power sources nowadays. MPC has emerged as a powerful design tool in this regard with its capabilities of handling various constraints and achieving optimal objectives. However, existing methods in this regard often involve computationally complex online constrained optimization, facing challenges to transition into real-world battery management systems. 

To overcome this issue, this paper proposed to exploit eMPC to enhance charging control, which inherits all the merits of MPC but enables highly efficient computation. 
Our design started with formulating an MPC charging control problem based on the NDC model. As the model is nonlinear, multi-segment linearization was developed to approximate the original MPC problem by a combination of multiple linear MPC problems. The solutions to the linear MPCs were computed offline and explicitly expressed as PWA functions, which made up an eMPC charging control algorithm. Contrasting the previous counterparts, this new algorithm is tremendously easy to code and fast to run online, potentially applicable to embedded computing hardware. The simulation and experimental results verified the effectiveness of the proposed design. Our future work will include cycle testing to assess how much the design can alleviate a LiB's aging process and incorporation of the temperature  factor into the design.

\balance
\bibliographystyle{IEEEtran}
\bibliography{TXT_TII-20-0598}

\begin{thebibliography}{10}
\providecommand{\url}[1]{#1}
\csname url@samestyle\endcsname
\providecommand{\newblock}{\relax}
\providecommand{\bibinfo}[2]{#2}
\providecommand{\BIBentrySTDinterwordspacing}{\spaceskip=0pt\relax}
\providecommand{\BIBentryALTinterwordstretchfactor}{4}
\providecommand{\BIBentryALTinterwordspacing}{\spaceskip=\fontdimen2\font plus
\BIBentryALTinterwordstretchfactor\fontdimen3\font minus
  \fontdimen4\font\relax}
\providecommand{\BIBforeignlanguage}[2]{{%
\expandafter\ifx\csname l@#1\endcsname\relax
\typeout{** WARNING: IEEEtran.bst: No hyphenation pattern has been}%
\typeout{** loaded for the language `#1'. Using the pattern for}%
\typeout{** the default language instead.}%
\else
\language=\csname l@#1\endcsname
\fi
#2}}
\providecommand{\BIBdecl}{\relax}
\BIBdecl

\bibitem{Wang:CSM:2018}
Y.~{Wang}, H.~{Fang}, L.~{Zhou}, and T.~{Wada}, ``Revisiting the
  state-of-charge estimation for lithium-ion batteries: A methodical
  investigation of the extended {K}alman filter approach,'' \emph{IEEE Control
  Systems Magazine}, vol.~37, no.~4, pp. 73--96, 2017.

\bibitem{hussein2011review}
A.~A.-H. Hussein and I.~Batarseh, ``A review of charging algorithms for nickel
  and lithium battery chargers,'' \emph{IEEE Transactions on Vehicular
  Technology}, vol.~60, no.~3, pp. 830--838, 2011.

\bibitem{cope1999art}
R.~C. Cope and Y.~Podrazhansky, ``The art of battery charging,'' in
  \emph{Fourteenth Annual Battery Conference on Applications and Advances.
  Proceedings of the Conference}.\hskip 1em plus 0.5em minus 0.4em\relax IEEE,
  1999, pp. 233--235.

\bibitem{Suthar:PCCP:2014}
B.~Suthar, V.~Ramadesigan, S.~De, R.~D. Braatz, and V.~R. Subramanian,
  ``Optimal charging profiles for mechanically constrained lithium-ion
  batteries,'' \emph{Phys. Chem. Chem. Phys.}, vol.~16, no.~1, pp. 277--287,
  2014.

\bibitem{Fang:JES:2018}
H.~Fang, C.~Depcik, and V.~Lvovich, ``Optimal pulse-modulated lithium-ion
  battery charging: Algorithms and simulation,'' \emph{Journal of Energy
  Storage}, vol.~15, pp. 359--367, 2018.

\bibitem{Hu:JPS:2013}
X.~Hu, S.~Li, H.~Peng, and F.~Sun, ``Charging time and loss optimization for
  {LiNMC} and {LiFePO4} batteries based on equivalent circuit models,''
  \emph{Journal of Power Sources}, vol. 239, pp. 449--457, 2013.

\bibitem{liu2016battery}
K.~Liu, K.~Li, Z.~Yang, C.~Zhang, and J.~Deng, ``Battery optimal charging
  strategy based on a coupled thermoelectric model,'' in \emph{2016 IEEE
  Congress on Evolutionary Computation}, 2016, pp. 5084--5091.

\bibitem{Perez:TVT:2017}
H.~E. {Perez}, X.~{Hu}, S.~{Dey}, and S.~J. {Moura}, ``Optimal charging of
  li-ion batteries with coupled electro-thermal-aging dynamics,'' \emph{IEEE
  Trans. Veh. Technol.}, vol.~66, no.~9, pp. 7761--7770, 2017.

\bibitem{liu2018charging}
K.~Liu, C.~Zou, K.~Li, and T.~Wik, ``Charging pattern optimization for
  lithium-ion batteries with an electrothermal-aging model,'' \emph{IEEE Trans.
  Ind. Informat.}, vol.~14, no.~12, pp. 5463--5474, 2018.

\bibitem{Fang:TCST:2017}
H.~Fang, Y.~Wang, and J.~Chen, ``Health-aware and user-involved battery
  charging management for electric vehicles: Linear quadratic strategies,''
  \emph{IEEE Trans. Control Syst. Technol.}, vol.~25, no.~3, pp. 911--923,
  2017.

\bibitem{Klein:ACC:2011}
R.~Klein, N.~A. Chaturvedi, J.~Christensen, J.~Ahmed, R.~Findeisen, and
  A.~Kojic, ``Optimal charging strategies in lithium-ion battery,'' in
  \emph{Proc. of American Control Conference}.\hskip 1em plus 0.5em minus
  0.4em\relax IEEE, 2011, pp. 382--387.

\bibitem{Liu:DSMC:2016}
J.~Liu, G.~Li, and H.~K. Fathy, ``A computationally efficient approach for
  optimizing lithium-ion battery charging,'' \emph{Journal of Dynamic Systems,
  Measurement, and Control}, vol. 138, no.~2, p. 021009, 2016.

\bibitem{Zou:Mechatronics:2018}
C.~Zou, C.~Manzie, and D.~Ne{\v{s}}i{\'c}, ``Model predictive control for
  lithium-ion battery optimal charging,'' \emph{IEEE/ASME Transactions on
  Mechatronics}, vol.~23, no.~2, pp. 947--957, 2018.

\bibitem{Torchio:ACC:2015}
M.~Torchio, N.~A. Wolff, D.~M. Raimondo, L.~Magni, U.~Krewer, R.~B. Gopaluni,
  J.~A. Paulson, and R.~D. Braatz, ``Real-time model predictive control for the
  optimal charging of a lithium-ion battery,'' in \emph{Proceedings of American
  Control Conference}.\hskip 1em plus 0.5em minus 0.4em\relax IEEE, 2015, pp.
  4536--4541.

\bibitem{Yan:Energies:2011}
J.~Yan, G.~Xu, H.~Qian, Y.~Xu, and Z.~Song, ``Model predictive control-based
  fast charging for vehicular batteries,'' \emph{Energies}, vol.~4, no.~8, pp.
  1178--1196, 2011.

\bibitem{Liu:JPS:2017}
K.~Liu, K.~Li, and C.~Zhang, ``Constrained generalized predictive control of
  battery charging process based on a coupled thermoelectric model,''
  \emph{Journal of Power Sources}, vol. 347, pp. 145--158, 2017.

\bibitem{Xavier:JPS:2015}
M.~A. Xavier and M.~S. Trimboli, ``Lithium-ion battery cell-level control using
  constrained model predictive control and equivalent circuit models,''
  \emph{Journal of Power Sources}, vol. 285, pp. 374--384, 2015.

\bibitem{Zou:Energy:2017}
C.~Zou, X.~Hu, Z.~Wei, and X.~Tang, ``Electrothermal dynamics-conscious
  lithium-ion battery cell-level charging management via state-monitored
  predictive control,'' \emph{Energy}, vol. 141, pp. 250--259, 2017.

\bibitem{Ouyang:TII:2018}
Q.~Ouyang, J.~Chen, J.~Zheng, and H.~Fang, ``Optimal multi-objective charging
  for lithium-ion battery packs: A hierarchical control approach,'' \emph{IEEE
  Transactions on Industrial Informatics}, 2018.

\bibitem{bemporad2002explicit}
A.~Bemporad, M.~Morari, V.~Dua, and E.~N. Pistikopoulos, ``The explicit linear
  quadratic regulator for constrained systems,'' \emph{Automatica}, vol.~38,
  no.~1, pp. 3--20, 2002.

\bibitem{bemporad2015multiparametric}
A.~Bemporad \emph{et~al.}, ``A multiparametric quadratic programming algorithm
  with polyhedral computations based on nonnegative least squares,'' \emph{IEEE
  Trans. Automat. Contr.}, vol.~60, no.~11, pp. 2892--2903, 2015.

\bibitem{Tian:IECON:2018}
N.~Tian, H.~Fang, and J.~Chen, ``A new nonlinear double-capacitor model for
  rechargeable batteries,'' in \emph{Proc. of the 44th Annual Conf. of the IEEE
  Industrial Electronics Society}, 2018, pp. 1613--1618.

\bibitem{woodford2013electrochemical}
W.~H. Woodford~IV, ``Electrochemical shock: Mechanical degradation of
  ion-intercalation materials,'' Ph.D. dissertation, Massachusetts Institute of
  Technology, 2013.

\bibitem{bandhauer2011critical}
T.~M. Bandhauer, S.~Garimella, and T.~F. Fuller, ``A critical review of thermal
  issues in lithium-ion batteries,'' \emph{Journal of the Electrochemical
  Society}, vol. 158, no.~3, pp. R1--R25, 2011.

\bibitem{pinson2013theory}
M.~B. Pinson and M.~Z. Bazant, ``Theory of {SEI} formation in rechargeable
  batteries: {C}apacity fade, accelerated aging and lifetime prediction,''
  \emph{J. Electrochem. Soc.}, vol. 160, no.~2, pp. A243--A250, 2013.

\bibitem{tian2019NDC}
N.~Tian, H.~Fang, C.~Jian, and Y.~Wang, ``Nonlinear double-capacitor model for
  rechargeable batteries: Modeling, identification and validation,'' \emph{IEEE
  Trans. Control Syst. Technol.}, to be published.

\bibitem{Borrelli:CUP:2017}
F.~Borrelli, A.~Bemporad, and M.~Morari, \emph{Predictive Control for Linear
  and Hybrid Systems}.\hskip 1em plus 0.5em minus 0.4em\relax Cambridge
  University Press, 2017.

\bibitem{bemporad2018model}
A.~Bemporad, N.~L. Ricker, and M.~Morari, \emph{Model {P}redictive {C}ontrol
  {T}oolbox {U}ser's {G}uide {R}2018b}.\hskip 1em plus 0.5em minus 0.4em\relax
  The {M}ath{W}orks, 2018.

\bibitem{tondel2003evaluation}
P.~T{\o}ndel, T.~A. Johansen, and A.~Bemporad, ``Evaluation of piecewise affine
  control via binary search tree,'' \emph{Automatica}, vol.~39, no.~5, pp.
  945--950, 2003.

\bibitem{gersnoviez2017high}
A.~Gersnoviez, M.~Brox, and I.~Baturone, ``High-speed and low-cost
  implementation of explicit model predictive controllers,'' \emph{IEEE Trans.
  Control Syst. Technol.}, vol.~27, no.~2, pp. 647--662, 2017.

\bibitem{fang2018nonlinear}
H.~Fang, N.~Tian, Y.~Wang, M.~Zhou, and M.~A. Haile, ``Nonlinear {B}ayesian
  estimation: From {K}alman filtering to a broader horizon,'' \emph{IEEE/CAA
  Journal of Automatica Sinica}, vol.~5, no.~2, pp. 401--417, 2018.

\end{thebibliography}

\end{document}